\documentclass[
    reprint,
    aps,
    amsmath,
    amssymb,
    superscriptaddress]{revtex4-2}
\usepackage{graphicx} % Include figure files
\usepackage{dcolumn}  % Align table columns on decimal point
\usepackage{bm}       % bold math
\usepackage{amsfonts}
\usepackage{dsfont}
\usepackage{amsmath}
\usepackage{url}
\usepackage{ulem}
\usepackage{color}

\begin{document}

\title{Enhancement of valley selective excitation by a linearly polarized two-color laser pulse}

\author{Arqum Hashmi}
\affiliation{%
Kansai Photon Science Institute, National Institutes for Quantum  Science and Technology (QST), Kyoto 619-0215, Japan }
            
\author{Shunsuke Yamada}
\affiliation{%
Kansai Photon Science Institute, National Institutes for Quantum Science and Technology (QST), Kyoto 619-0215, Japan }

\author{Kazuhiro Yabana}
\affiliation{%
Kansai Photon Science Institute, National Institutes for Quantum  Science and Technology (QST), Kyoto 619-0215, Japan }
\affiliation{%
Center for Computational Sciences, University of Tsukuba, Tsukuba 305-8577, Japan}

\author{Tomohito Otobe}
\email[]{otobe.tomohito@qst.go.jp}
\affiliation{%
Kansai Photon Science Institute, National Institutes for Quantum  Science and Technology (QST), Kyoto 619-0215, Japan }
%\affiliation{%
%Photon Science Center, The University of Tokyo, Hongo Bunkyoku,
%Tokyo 111-8656, Japan}            

\date{\today}% It is always \today, today,

\begin{abstract}
%One of the most interesting features of transition metal dichalcogenides is the electron’s extra degree of freedom, the valley pseudospin. This extra degree of freedom has potential in the field of valleytronics. 
Here we proposed the valley selective excitations via a two-color (\ensuremath{\omega} + \ensuremath{2\omega}) laser field, made by superimposing two linearly polarized 
pulses at frequencies \ensuremath{\omega} and \ensuremath{2\omega}. 
We have studied the intensity ratio between a few-cycle 
pulse of \ensuremath{\omega} and \ensuremath{2\omega} laser, and 
its enhancement factor by employing the time-dependent first-principle calculations. 
%The valley polarization predominantly depends on the excitation dynamics by \ensuremath{\omega} + \ensuremath{2\omega}. 
The valley polarization depends on the carrier envelope phases (CEPs) of pulses and the intensity ratio $I_{\omega}/I_{2\omega}$.
We found that the two-color field enhances the valley polarization  as much as 1.2 times larger than the single-color pulse. 
The maximum valley asymmetry is achieved for the intensity ratio $I_{\omega}/I_{2\omega}$ of 36 
with the relative CEP of \ensuremath{\pi}.
%which is much higher than the case of molecules where the ratio $I_{\omega}/I_{2\omega}$ is 9.
In our previous work, we found that the asymmetric vector potential induces 
the valley polarization (Phys. Rev. B 105,115403 (2022)). 
In this work, we find that the asymmetry of the electric field modulates the valley polarization.
%in the case of \ensuremath{\omega} + \ensuremath{2\omega} field. 
%Nonetheless, the symmetry of the vector potential is the critical parameter 
%and that of the electric field plays a minor role. 
%The valley polarization has a peak when the relative phase between the \ensuremath{\omega} and \ensuremath{2\omega} pulses is \ensuremath{\pi}.
Our two-color scheme offers a new path toward the optical control of valley pseudospins.

\end{abstract}

\maketitle

\section{INTRODUCTION }

Electrons in two-dimensional (2D) hexagonal lattices have an extra
degree of freedom in addition to charge and spin named valley pseudospin
\citep{xiao2012coupled,xu2014spin}. Valleys are local minima in the band structure corresponding to different crystal momenta
that are located at the $K$/$K^{'}$ points of the Brillouin zone
(BZ) \citep{wang2012electronics,schaibley2016valleytronics}. In the
field of valleytronics, 2D layer materials such as graphene \citep{xiao2007valley,shimazaki2015generation}
and transitional metal dichalcogenides (TMDs) \citep{sie2018valley,cao2012valley,jones2013optical}
are receiving extensive research efforts. Graphene presents the remarkable
feature of Dirac fermions \citep{geim2009graphene,neto2009electronic}.
However, the honeycomb structure with equivalent A and B sublattice
enforces zero Berry curvature which makes it not an ideal candidate
for valley contrasting properties \citep{zhang2005experimental}.
As opposed to graphene, TMDs monolayers are of particular interest
for practical valleytronics applications because of their broken inversion
symmetry and strong spin-orbit coupling (SOC).

Lifting the valley degeneracy to create valley polarization has become a central theme in valleytronics \citep{yuan2014generation,vitale2018valleytronics}.
Several methods have been proposed to achieve transient valley polarization
such as applying a magnetic field \citep{macneill2015breaking}, optical
Stark effect \citep{kim2014ultrafast}, and proximity effects generated
by magnetic substrates \citep{zhong2017van}. Due to the several practical
limitations of the magnetic field, the optical excitations remain
a popular way where selective excitation at $K$ or $K^{'}$ can be
achieved using a weak circularly polarized field resonant with the bandgap of the material\citep{cao2012valley,zeng2012valley,mak2012control}.
Depending on its helicity, the field couples to either $K$/$K^{'}$valleys.

The valley-dependent optical selection rules suggest that linearly
polarized light couples equal to both valleys. It is widely
accepted that valley polarization with linearly polarized fields is
impossible \citep{schaibley2016valleytronics,yao2008valley}. In
contrast, it is expected that  a few-cycle single-color pulse with the controlled
carrier-envelope phase (CEP) can break this situation
\citep{jimenez2021sub,hashmi2022valley}. Moreover, ultrashort laser
pulses offer ultrafast control of electron dynamics \citep{langer2018lightwave,kumar2021ultrafast}.
The possibility of generating the valley polarization with linearly
polarized pulses offers an alternative route to valleytronics in graphene
and TMDs materials \citep{jimenez2021sub,kelardeh2022ultrashort}.
The advantage of linearly polarized pulses is to avoid
reliance on resonant pump-probe spectroscopy that is used in TMDs
monolayers to break the symmetry between the $K$ and $K^{'}$ valleys.

Electron dynamics under a strong laser field can be described by 
solving the time-dependent Kohn-Sham (TDKS) equation in 
real-time referred to as time-dependent density functional theory (TDDFT) \citep{theilhaber1992ab, yabana1996time}. 
TDDFT is not only been used to describe the linear response in the frequency domain \citep{yabana2006real,laurent2013td} but also has been very successful 
to describe the nonlinear and nonperturbative 
dynamics of electrons by intense
ultrashort laser pulses \citep{castro2004excited,nobusada2004high,otobe2008first}. 
The most powerful aspect of TDDFT is its capability to describe 
electron dynamics under intense laser fields without any empirical 
parameters. In condensed matter, it is well known that SOC modifies the band structure of solids.  Strong SOC not only modifies the band 
curvature but also changes the bandgap of the material by 
lifting the spin degeneracy which ultimately
can affect the excitation of carriers. Hence the inclusion of the 
SOC in TDDFT is essential not only to describe the excitation 
dynamic under a strong laser field but also important to accurately describe 
the phenomenon such as spintronics \citep{vzutic2004spintronics} or valleytronics \citep{vitale2018valleytronics}. 

In our previous study, we found that the asymmetric laser field with mono-cycle
 laser pulse enables the valley polarization by the real-time TDDFT approach with SOC \citep{hashmi2022valley}. 
A two-color laser can also control the asymmetric excitation \citep{Ohmura2014,kaziannis2014interaction, higuchi2017light,jimenez2020lightwave}. 
In this work, we study the enhancement of valley polarization in WSe$_2$ 
monolayer via a two-color field.
It is possible to create an asymmetric laser field by mixing a fundamental with 
frequency \ensuremath{\omega} and its second harmonic \ensuremath{2\omega}.
The two pulses intensity ratio and the relative CEP control the valley polarization. We also compare the valley polarization results
of \ensuremath{\omega} + \ensuremath{2\omega} scheme results
with the single color (\ensuremath{\omega}) pulse. We found that
the valley polarization via two-color-control exceeds the single-color
scheme by 1.2 times as the \ensuremath{\omega} + \ensuremath{2\omega}
field exhibits more asymmetry in its temporal shape. Our results reveal
a convenient new path toward the optical control of valley pseudospins.

\section{THEORETICAL FORMALISM}

Using the velocity gauge \citep{bertsch2000real}, we describe the
time evolution of electron orbitals in WSe$_{2}$ monolayer under
the pulsed electric field by solving the TDKS equation for Bloch orbitals $u_{b,{\bf k}}({\bf r},t)$ (which
is a two-component spinor. $b$ is the band index and ${\bf k}$ is
the 2D crystal momentum of the thin layer) as,

\begin{equation}
\begin{split}i\hbar\frac{\partial}{\partial t}u_{b,{\bf k}}({\bf r},t)=\Big[\frac{1}{2m}{\left(-i\hbar\nabla+\hbar{\bf k}+\frac{e}{c}{\bf A}^{{\rm (t)}}(t)\right)}^{2}\\
-e\varphi({\bf r},t)+\hat{v}_{{\rm NL}}^{{{\bf k}+\frac{e}{\hbar c}{\bf A}^{{\rm (t)}}(t)}}+{v}_{{\rm xc}}({\bf r},t)\Big]u_{b,{\bf k}}({\bf r},t),
\end{split}
\label{1}
\end{equation}
where the scalar potential $\varphi({\bf r},t)$ includes the Hartree
potential from the electrons plus the local part of the ionic pseudopotentials
and we have defined $\hat{v}_{{\rm NL}}^{{\bf k}}\equiv e^{-i{\bf k}\cdot{\bf r}}\hat{v}_{{\rm NL}}e^{i{\bf k}\cdot{\bf r}}$.
Here, $\hat{v}_{{\rm NL}}$ and ${v}_{{\rm xc}}({\bf r},t)$ are the
nonlocal part of the ionic pseudopotentials and exchange-correlation
potential, respectively.

TDKS equation combined with the Maxwell equations can describe the
light propagation in thin layers as well. For this purpose, we assume
the spatially uniform macroscopic electric field inside the layer
(in our case monolayer is in the $xy$ plane and the light pulse propagates
along the $z$ axis) \citep{yamada2018time,yamada2021determining}.
By using the Maxwell equations, the propagation of the macroscopic
electromagnetic fields in the form of the vector potential ${\bf A}(z,t)$
is described as, 
\begin{equation}
\left(\frac{1}{c^{2}}\ \frac{{\partial}^{2}}{{\partial t}^{2}}-\ \frac{{\partial}^{2}}{{\partial z}^{2}}\right){\bf A}\left(z,t\right)=\ \frac{4\pi}{c}{\bf J}(z,t),
\label{2}
\end{equation}
where ${\bf J}(z,t)$ is the macroscopic current density of the thin
layer.

Under the assumption of zero thickness and spatially uniform electric
field inside the monolayer, we can approximate the macroscopic electric
current density in Eq. (\ref{2}) as 
\begin{equation}
{\bf J}\left(z,t\right)\approx\delta(z){\bf J}_{{\rm 2D}}(t),
\label{3}
\end{equation}
where ${\bf J}_{{\rm 2D}}(t)$ is 2D current density (current per
unit area) of the thin layer.

By treating it as a boundary value problem where reflected (transmitted)
fields can be determined by the connection conditions at $z=0$, we
can get the ${\bf J}_{{\rm 2D}}(t)$ as follows,

\begin{equation}
\begin{split}{\bf J}_{{\rm 2D}}(t)=-\frac{e}{m}\int dz\int_{\Omega}\frac{dxdy}{\Omega}\sum_{b,{\bf k}}^{{\rm occ}}u_{b,{\bf k}}^{\dagger}({\bf r},t)\\
\times\left[-i\hbar\nabla+\hbar{\bf k}+\frac{e}{c}{\bf A}^{{\rm (t)}}(t)+\frac{m}{i\hbar}\left[{\bf r},\hat{v}_{{\rm NL}}^{{{\bf k}+\frac{e}{\hbar c}{\bf A}^{{\rm (t)}}(t)}}\right]\right]u_{b,{\bf k}}({\bf r},t),
\end{split}
\label{4}
\end{equation}
where $\Omega$ is the area of the 2D unit cell and the sum is taken
over the occupied orbitals in the ground state. The detail of our
formalism is explained in the references \citep{hashmi2022valley,yamada2018time}.

The method explained above is implemented in the TDDFT package named
Scalable Ab initio Light-Matter simulator for Optics and Nanoscience
(SALMON). Full details of the SALMON code and its implementation are
described elsewhere\ \citep{noda2019salmon,Salmon:Online}. 

The lattice parameter of WSe$_{2}$ monolayer is chosen to be $3.32$~{Å}.
The vacuum distance of 20 Å was employed in the direction normal to
the interface. The exchange-correlation potential utilizes the adiabatic
local-density approximation with Perdew-Zunger functional\ \citep{perdew1992accurate}.
We use a spin non-collinear treatment for the exchange-correlation
potential\citep{von1972local,oda1998fully}. The dynamics of the 12 valence electrons
for one W atom, and 12 valence electrons for two Se atoms are treated explicitly
while the effects of the core electrons are considered through norm-conserving
pseudopotentials \ \citep{morrison1993nonlocal}. The total energy
convergence criterion was set to 10$^{-8}$ eV. For such calculations,
the spatial grid sizes and k-points are optimized. The converged results were obtained with a fine
r-grid and k-grid of 0.21~{Å} and 15 \texttimes{} 15 \texttimes{}
1 k-mesh respectively. We use the vector potential for the fundamental laser pulse and its second harmonics pulses with the following waveform, 
\begin{equation}
\begin{split}A\left(t\right)=-\frac{c}{\omega}f\left(t\right)
\Bigg[E_{\omega}{\cos}\left\{ \omega\left(t-\frac{T_{P}}{2}\right)+ \varphi_{\omega}\right\} \\
+\frac{1}{2}E_{2\omega}{\cos}\left\{ 2\omega\left(t-\frac{T_{P}}{2}\right)+ \varphi_{2\omega}\right\}\Bigg]\end{split}
\label{5}
\end{equation}
where, $E_{\omega}$ and $E_{2\omega}$ are the peak electric field amplitudes
while $\varphi_{\omega}$ and $\varphi_{2\omega}$ are CEP of the fundamental
pulse and its second harmonics respectively. $\omega$ is the carrier
frequency and $T_{P}$ is the total pulse duration. The pulse envelope
function is of $\cos^{4}$ shape for the vector potential given as
\begin{equation}
\begin{aligned}f(t)= & \begin{cases}
\cos^{4}\left(\pi\frac{t-T_{P}/2}{T_{P}}\right) & 0\leq t\leq T_{P}\\
0 & \mathrm{otherwise}
\end{cases}\;.\end{aligned}
\label{6}
\end{equation}
We use the linearly polarized pulses with 0.4~eV frequency for the fundamental and its second harmonic (0.8~eV). Both pulse lengths are set to $T_{P}=10$~fs and the small time
step size of 5$\mathrm{\times}$10$^{-4}$~fs is used for stable
calculations.

\section{RESULTS AND DISCUSSION}

\begin{figure*}
\centering \includegraphics[scale=0.6]{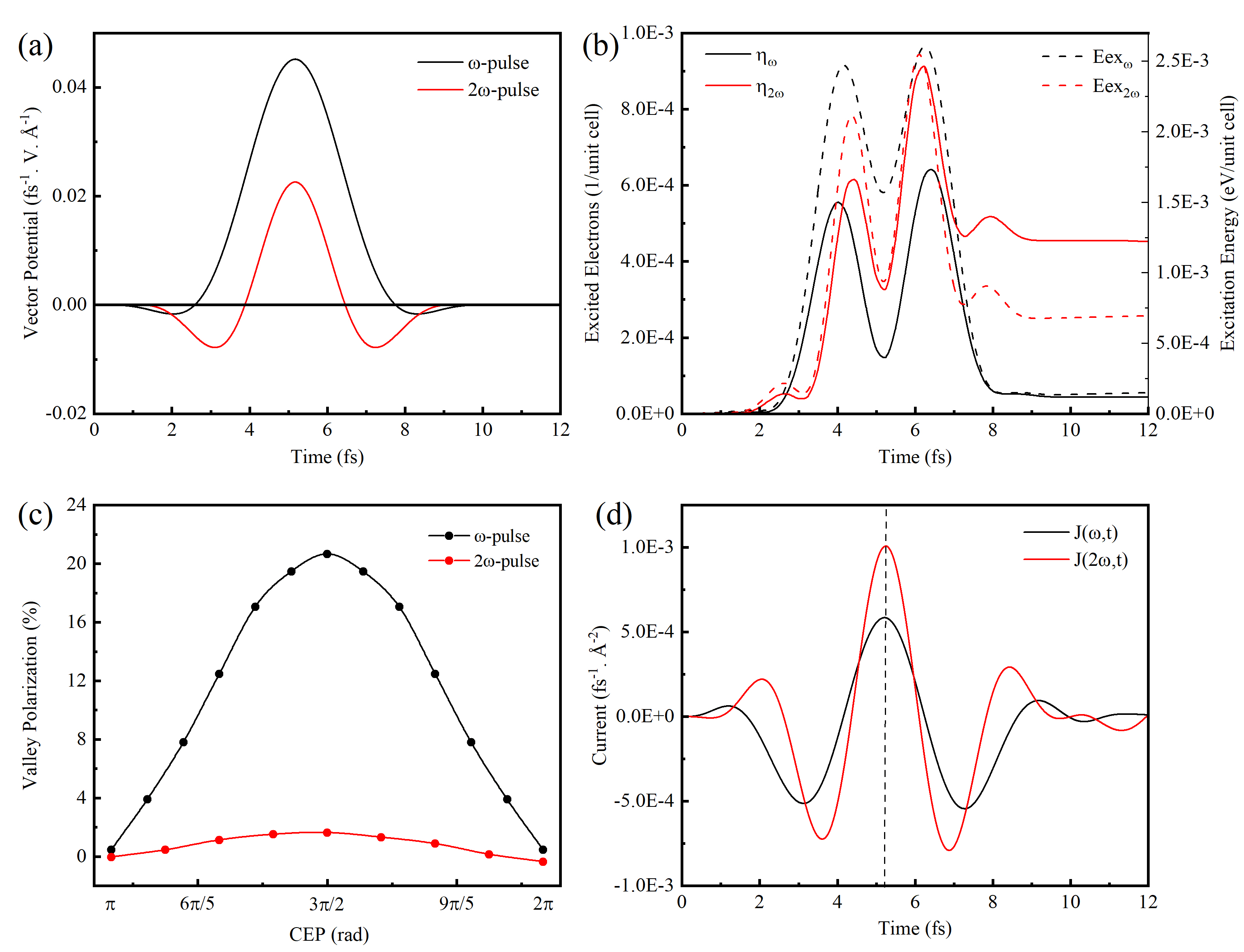} \caption{ Comparison of two single-color laser fields with the frequencies of \ensuremath{\omega} and 2\ensuremath{\omega}. Both pulses have the intensity of 10$^{10}$ W/cm$^{2}$. (a) The vector potential  (b) The temporal development of excitation energies (Eex) and excited electrons (\ensuremath{\eta}) (c) Valley polarization as a function of CEP and (d) The temporal evolution of photocurrents. The CEP in panels (a), (b), and (d) is \ensuremath{3\pi}/2. }
\label{fig:1} 
\end{figure*}
For the molecules, the asymmetric behavior with the intensity ratio \ensuremath{I_{\omega}/I_{2\omega}} around four have been used with the photon energy far below the ionization potential \citep{Ohmura2014, kaziannis2014interaction}. 
Here, $I_{\omega}$ and $I_{2\omega}$ are the intensity of \ensuremath{\omega} and \ensuremath{2\omega} pulses respectively. Since we assume the frequencies comparable to the bandgap of WSe$_2$, the electronic response is significantly different for \ensuremath{\omega} and \ensuremath{2\omega}. Therefore, we have to clarify the electron dynamics under each color field before proceeding with the two-color field.
%First, we individually examine the excitation dynamics and its effect on valley polarization by \ensuremath{\omega} and \ensuremath{2\omega} fields.

Fig. \ref{fig:1}(a) shows the vector potential of both pulses.
The field intensity of I = 10$^{10}$ W/cm$^{2}$ is used for \ensuremath{\omega}  
and \ensuremath{2\omega} pulses, the laser intensity is chosen to induce the non-linear dynamics but be below the damage threshold for 
monolayer TMDs \citep{liu2017high,yoshikawa2019interband}. 
In our previous study, we have shown that the polarization parallel to $\mathit{\Gamma}$-$K$ experiences different 
band curvature and hence induces a high degree of valley polarization that can be controlled by CEP \citep{hashmi2022valley}. 
Thus the direction of the polarization of the 
incident light is considered along the $\mathit{\Gamma}$-$K$. 
Fig. \ref{fig:1}(b) displays the excitation dynamics 
regarding excitation energy and excited electrons. In the case of \ensuremath{\omega}-field, the excitation energy 
and the excited electrons are noticeable during the pulse irradiation, and the electronic state returns to its ground state when the pulse ends. 
The energy per electron is about 8-photon (3.16~eV/electron), which is higher than the expected lowest excitation path (4-photon). 
%This non-linearity suggests a non-perturbative effect, such as the tunneling mechanism for electrons. 
On the other hand, in the case of \ensuremath{2\omega} field, more electrons are excited as compared to \ensuremath{\omega} 
and the excited electrons remain 
finite after the pulse ends. The absorbed energy per electron is about 1.52~eV/electron, which indicates the 
lowest order of the photo-absorption, the two-photon absorption, is dominant. 
The Keldysh parameter  is around 2 for $\omega$-field and 4 for $2\omega$-field.
These results suggest that the excitation process under \ensuremath{\omega}-field is highly non-linear than $2\omega$, and is in the intermediate stage between multi-photon and tunneling processes.

Fig. \ref{fig:1}(c) shows the comparison of the valley polarization by \ensuremath{\omega} and \ensuremath{2\omega} fields. 
The valley polarization is defined as, 
\begin{equation}
Polarization\mathrm{=\ }
\frac{{\rho}_{K^{'}} - {\rho}_{K}}{\frac{1}{2}\left(\rho_{K^{'}} + {\rho}_{K}\right)},
\label{7}
\end{equation}
where~${\rho}_{K}({\rho}_{K^{'}})$ are electron densities, obtained
by integrating the conduction band electron population around $K$($K^{'}$) point. 
Strong valley polarization is observed by 
\ensuremath{\omega}-pulse while the \ensuremath{2\omega}-field 
does not show any significant valley polarization. 
%These results suggest that the non-perturbative process induces strong valley
%polarization while the multiphotonic absorption destroys 
%the valley polarization mechanism. 
%Hence, the valley polarization is mainly 
%determined by the excitation dynamics. 
The photo-current indicates the sensitivity of the system for each frequency, shown in Fig. \ref{fig:1}(d). 
The current ratio for \ensuremath{2\omega} is roughly twice of \ensuremath{\omega}-pulse. 
The large difference in the photo-current between \ensuremath{2\omega} and \ensuremath{\omega}-pulses 
indicates that we have to find a good ratio $I_{\omega}/I_{2\omega}$ to enhance the asymmetric valley polarization.
%In the vacuum, the ratio $E_{\omega}/E_{2\omega}=3$ shows the most asymmetric field. For atoms and molecules, 
%this ratio is also the best because the polarizability is not so different for \ensuremath{\omega} 
%and \ensuremath{2\omega} in many cases. In WSe$_2$, electrons respond two times more intensely for 2omega than omega. 
%This result implies that the best asymmetric dynamics is induced by the ratio of $E_{\omega}/E_{2\omega}=6$. 
%The ratio of field $E_{\omega}/E_{2\omega}=6$ 
%corresponds to the intensity ratio of  $I_{\omega}/I_{2\omega}=36$, %where $I_{\omega}$ 
%and $I_{2\omega}$ are the intensity of \ensuremath{\omega} and %\ensuremath{2\omega} pulses respectively.  

\begin{figure*}
\centering \includegraphics[scale=0.5]{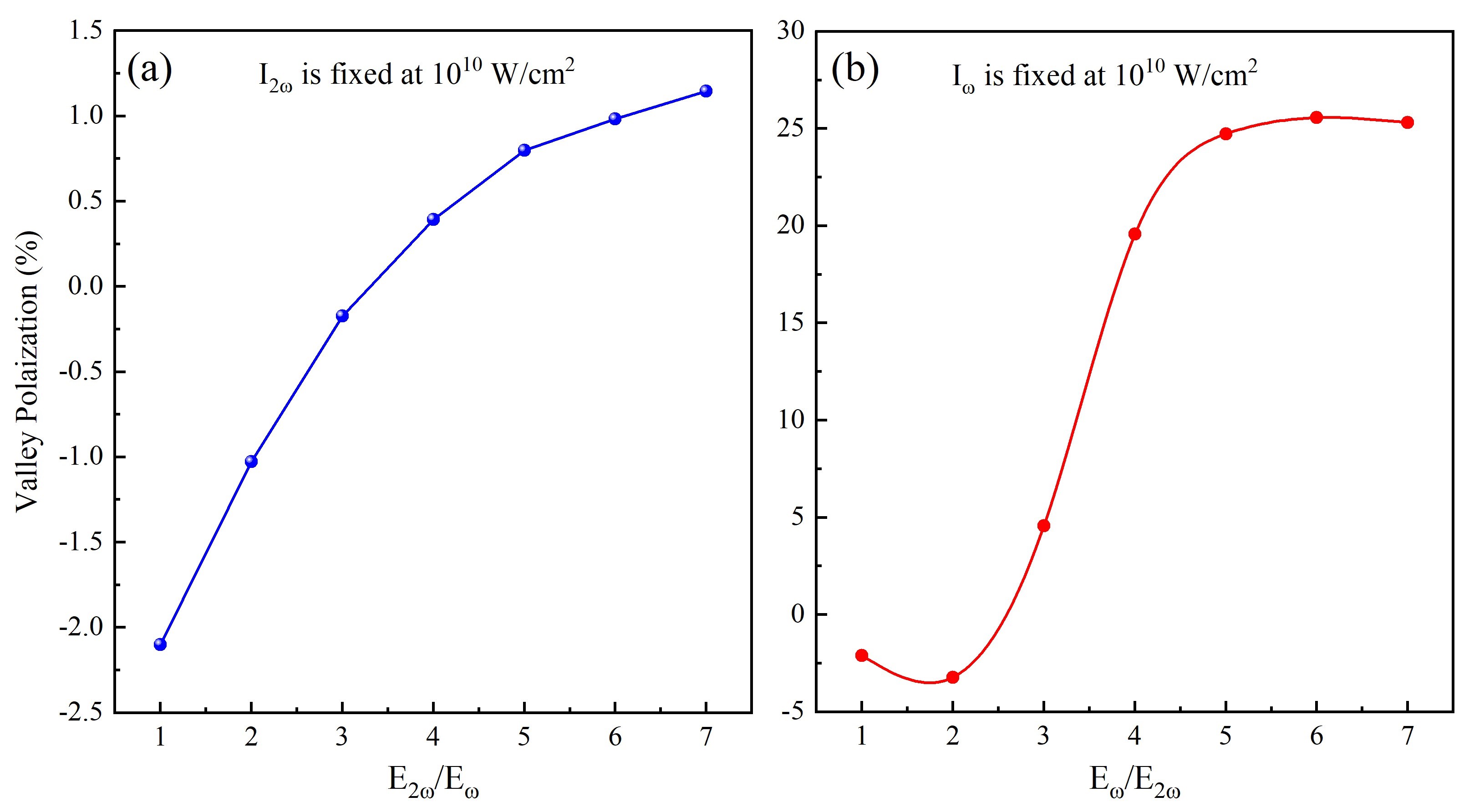} \caption{ Valley polarization of the two-color field (\ensuremath{\omega} + \ensuremath{2\omega}) field as the function of the ratio between the electric field strengths. (a) The field strength of \ensuremath{\omega} is varied while 2\ensuremath{\omega} is fixed. The CEP is fixed at \ensuremath{\varphi}$_{\omega}$ = \ensuremath{\pi}/2 and   \ensuremath{\varphi}$_{2\omega}$ = \ensuremath{3\pi}/2 (b) The field strength of \ensuremath{2\omega} is varied while \ensuremath{\omega} is fixed. The CEP is fixed at \ensuremath{\varphi}$_{\omega}$ = \ensuremath{3\pi}/2 and \ensuremath{\varphi}$_{2\omega}$ = \ensuremath{\pi}/2. The CEPs are chosen to have a positive peak of valley polarization.}
\label{fig:2} 
\end{figure*}

%\textcolor{red}{Here one should note that to create the asymmetric field by a single
%color pulse (\ensuremath{\omega}), the number of controllable laser
%parameter is just the CEP. In contrast, not
%only the CEP of the two laser fields but also the ratio of the field
%can be controlled in the Bichromatic \ensuremath{\omega}+2\ensuremath{\omega}
%pulses. To determine the ratio of the electric field strengths for
%the two fields that can induce maximum asymmetry on the WSe$_{2}$
%monolayer, 
%we have examined the field ratio between \ensuremath{\omega} and \ensuremath{2\omega}.} 
 
According to our previous work\citep{hashmi2022valley}, the valley polarization with 
the mono-cycle pulse shows the peak with the phase of \ensuremath{\varphi}$=\pi/2$ and $3\pi/2$.
Regarding the relative phase (\ensuremath{\varphi}$_{rel}$), 
the intense asymmetric response of molecules has been reported with the  
$0$ and $\pi$\citep{Ohmura2004}.
Fig. \ref{fig:2}(a) shows the valley polarization by changing 
the $E_{\omega}$ field while 
the  $E_{2\omega}$ is fixed at 10$^{10}$ W/cm$^{2}$, that is intense $2\omega$ pulse case.
The CEPs are set as \ensuremath{\varphi}$_{\omega}=\pi/2$ and \ensuremath{\varphi}$_{2\omega}=3\pi/2$ to 
induce the best valley polarization.
%The maximum value of valley polarization is observed 
%when \ensuremath{\varphi}= \textpm \ensuremath{\pi}\ensuremath{\slash}2 (-\ensuremath{\pi}/2 = \ensuremath{3\pi}/2, %\ensuremath{\pi}/2) for \ensuremath{\omega} and \ensuremath{2\omega},
%thus the set of \ensuremath{\varphi}$_{rel}$ that exhibit the best value of valley polarization is selected in Fig. \ref{fig:2}.
Intense \ensuremath{2\omega} 
results in two-photon absorption which leads the 
$K$ and $K^{'}$ valleys equally excited. 
Hence no valley asymmetry is present in this case.  

Fig. \ref{fig:2}(b) shows the valley 
polarization by changing the $E_{2\omega}$ with intense $E_{\omega}$. 
The valley polarization increases as the $E_{2\omega}$ decreases which distinctly indicates the change in the 
excitation process from the two-photon absorption to tunneling. 
%An increase in the valley asymmetry with the decrease 
%in $E_{2\omega}$ indicates that the two-photon absorption decreases.
Fig. \ref{fig:2} confirms that the maximum valley polarization 
is observed for the $E_{\omega}/E_{2\omega}$ ratio of 6. 
The ratio of field $E_{\omega}/E_{2\omega}=6$ 
corresponds to the intensity ratio of  $I_{\omega}/I_{2\omega}=36$.
It should be noted that the ratio $I_{\omega}/I_{2\omega}=4$, 
which is the ratio used in the molecule case, does not work (Fig.~\ref{fig:2}(b)).
For further calculations, we have fixed the laser intensity of the fundamental pulse to $I_{\omega}$  = 1$\mathrm{\times}$10$^{10}$ W/cm$^{2}$ 
while the superimposed \ensuremath{\omega} + \ensuremath{2\omega} 
optimized ratio $I_{\omega}/I_{2\omega}$ is fixed to 36.

\begin{figure*}
\centering\includegraphics[scale=0.55]{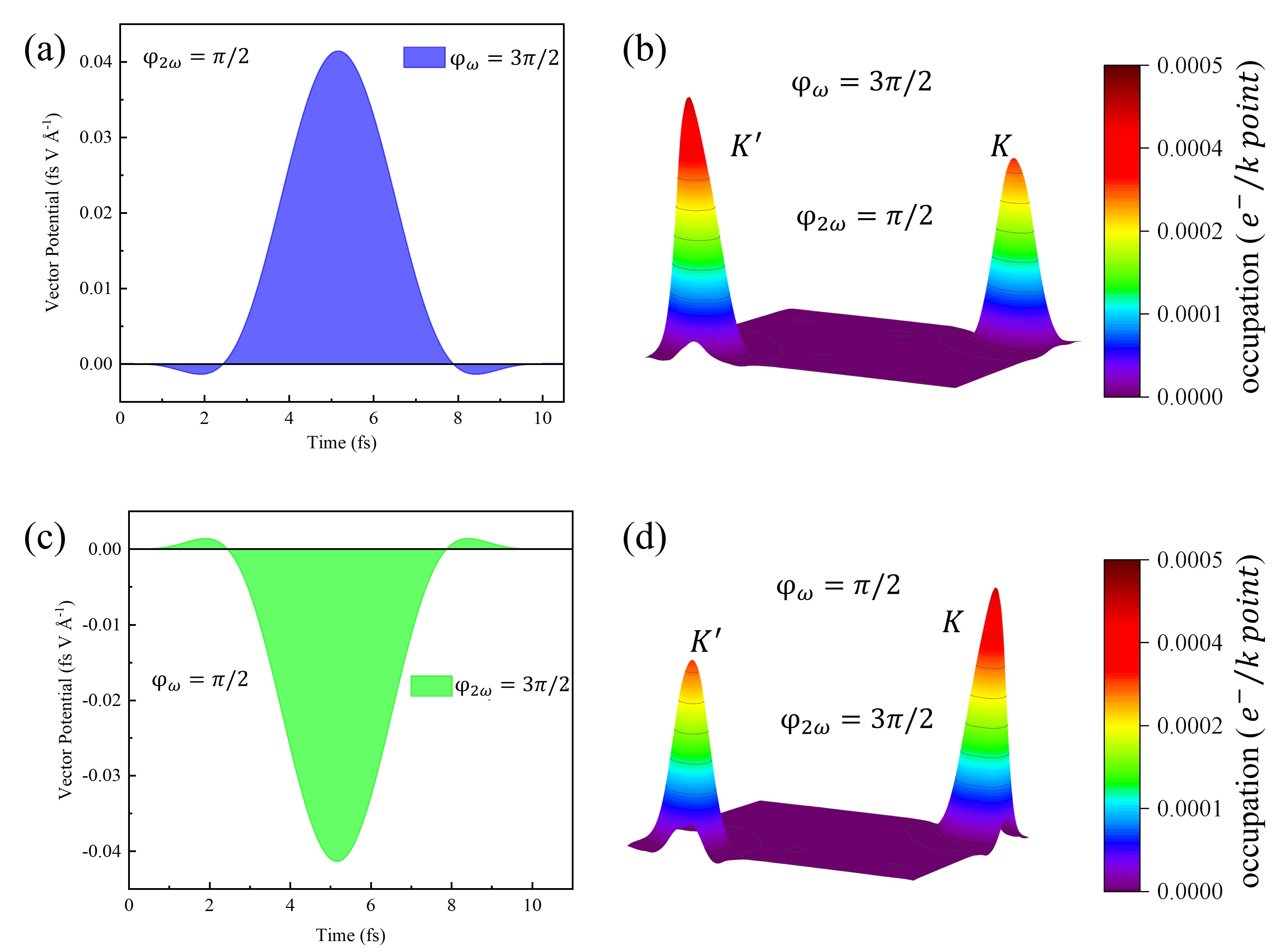} \caption{(a) Vector potential of the \ensuremath{\omega} + \ensuremath{2\omega} field for \ensuremath{\varphi}$_{\omega}$  \ensuremath{3\pi}/2 and \ensuremath{\varphi}$_{2\omega}$ \ensuremath{\pi}/2. (b) Conduction band electron populations for the vector potential in (a). The (c) and (d) are the same as (a) and (b) but for \ensuremath{\varphi}$_{\omega}$ = \ensuremath{\pi}/2, \ensuremath{\varphi}$_{2\omega}$ = \ensuremath{3\pi}/2. The electron population is summed over the entire conduction band at the end of the pulse. Two waveforms and their corresponding k-resolved populations are chosen because of their maximum asymmetry and the switching of the valley from $K^{'}$ to $K$.}
\label{fig:3} 
\end{figure*}

%\textcolor{red}{As we described above in this scheme, the other main factor is the
%relative phase \ensuremath{\varphi}$_{rel}$ between two pulses.}
%The most important factor is the \ensuremath{\varphi}$_{rel}$ to tune the valley polarization with a fixed intensity ratio. 
%The \ensuremath{\omega}-pulse mainly derived the asymmetry because it has a much higher field 
%strength than \ensuremath{2\omega}. 
%The potential imbalance changes with respect to $\left|A_{+}\right|$ 
%and $\left|A_{-}\right|$ as we vary the \ensuremath{\varphi}$_{\omega}$. 
%Here, $\left|A_{+}\right|$ and $\left|A_{-}\right|$ are the peak value in positive and negative directions respectively.
%The maximum asymmetry is achieved when both colors are separated by \ensuremath{\varphi}$_{rel} = \pi$ as shown in Fig. \ref{fig:3}(a).
%\textcolor{red}{When both colors are separated by \ensuremath{\varphi}$_{rel}$= \ensuremath{\pi}\ensuremath{\slash}2, A(t) is more asymmetric as compared to \ensuremath{\varphi}$_{rel}$
%= 0.
%The maximum asymmetry is achieved when both 
%colors are separated by \ensuremath{\varphi}$_{rel} = \pi$ that is shown in Fig. \ref{fig:3},
%thus we expected to observe 
%maximum asymmetry in carrier density with respect to $K$ and $K^{'}$  points in the BZ.}
Next, we investigate the distribution 
of excited state electrons in the reciprocal space. 
The excited electron population is defined as, 
\begin{equation}
\rho_{\bf {k}}(t)=\sum_{c,v}\left|\int_{\Omega}d^{3}r\,u_{v,{\bf k}}^{\ast}({\bf r},t)\,u_{c,{\bf k}+\frac{e}{\hbar c}{\bf A}^{{\rm (t)}}(t)}^{{\rm GS}}({\bf r})\right|^{2},
\label{8}
\end{equation}
where $v$ and $c$ are the indices for the valence and conduction bands, respectively, and 
$u_{b,{\bf k}}^{{\rm GS}}({\bf r})=u_{b,{\bf k}}({\bf r},t=0)$ 
is the Bloch orbital in the ground state.

Fig. \ref{fig:3}(b) shows the maximum population asymmetry 
at $K$ and $K^{'}$ valley corresponds to vector potential 
shown in Fig. \ref{fig:3}(a). One can see that positive 
vector potential favors the $K^{'}$ valley. 
As the \ensuremath{\varphi}$_{\omega}$ changes 
from \ensuremath{3\pi}/2 to \ensuremath{\pi}/2 the waveform of the vector potential changes 
from positive to negative as shown in Fig. \ref{fig:3}(c).
Reversing the \ensuremath{\varphi} of 
\ensuremath{\omega}-pulse (-\ensuremath{\pi}/2 (\ensuremath{3\pi}/2) → \ensuremath{\pi}/2)
switch the valley asymmetry from $K^{'}$ to $K$ valley 
as shown in Fig. \ref{fig:3}(d) suggests a robust mechanism for valley selection.

\begin{figure*}
\centering \includegraphics[scale=0.55]{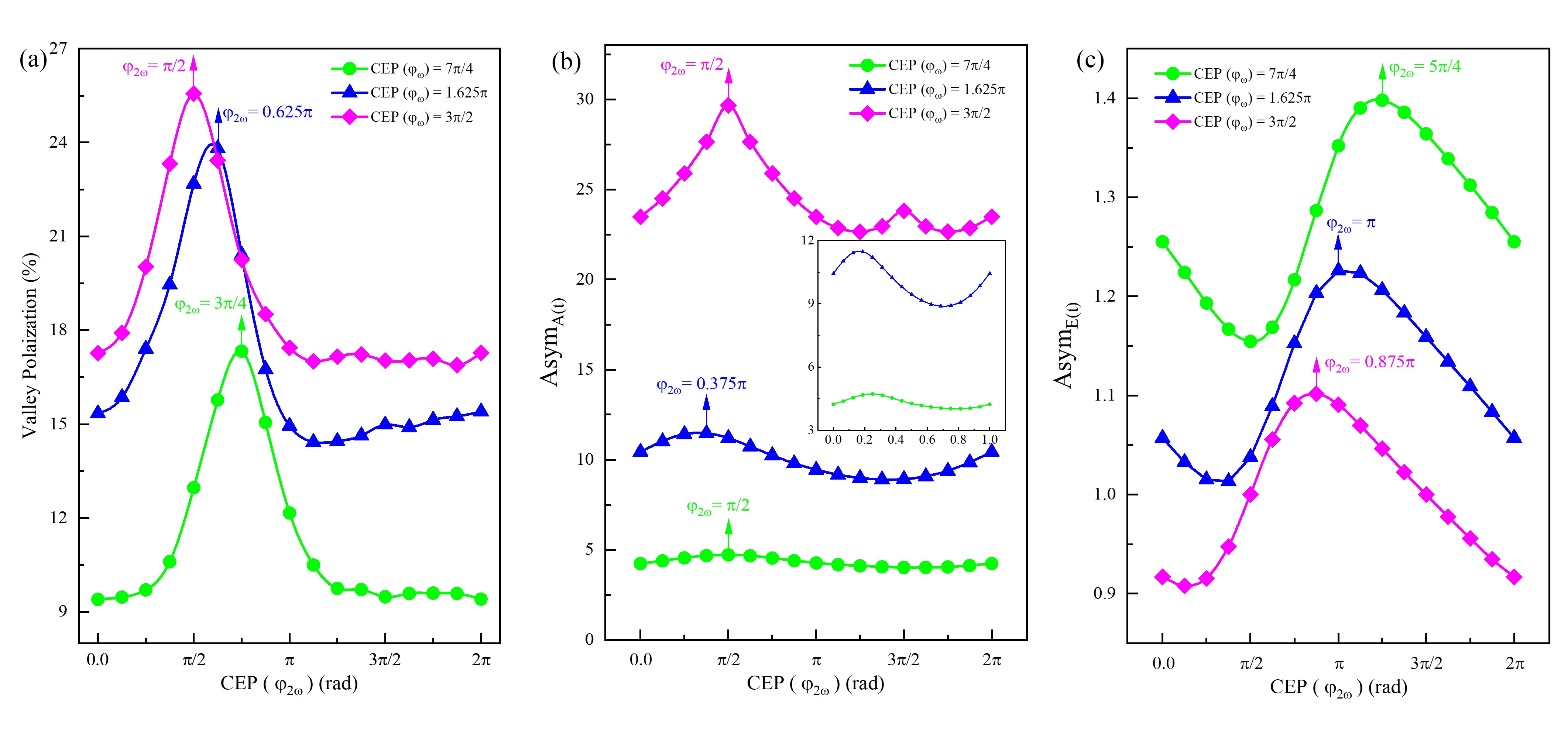} \caption{(a) Valley polarization (b) $Asym_{A(t)}$ and (c) $Asym_{E(t)}$ as a function of \ensuremath{\varphi}$_{2\omega}$ at specific CEP of \ensuremath{\omega}. The intensity of the I$_{\omega+2\omega}$ = 0.01 TWcm$^{-2}$ while the $I_{\omega}/I_{2\omega}$ = 36.}
\label{fig:4} 
\end{figure*}

Fig. \ref{fig:4}(a) shows the valley polarization as a function of the \ensuremath{\varphi}$_{2\omega}$ at numerous CEP of \ensuremath{\varphi}$_{\omega}$. 
As described above, no asymmetry is observed 
at \ensuremath{\varphi}$_{\omega}$= 0. Consequently, 
the valley polarization is zero regardless of 
the \ensuremath{\varphi}$_{2\omega}$. The valley polarization increases as we change the \ensuremath{\varphi}$_{\omega}$. 
The valley polarization has the order of
\ensuremath{\varphi}$_{\omega}$ = \ensuremath{3\pi}/2 $\mathrm{>}$ \ensuremath{\varphi}$_{\omega}$ = \ensuremath{1.625\pi} $\mathrm{>}$
\ensuremath{\varphi}$_{\omega}$ = \ensuremath{7\pi}/4 and the maximum
value is as high as 26\%.  
Valley polarization 
via two-color-control exceeds the single-color scheme by 
1.2 times as the \ensuremath{\omega} + \ensuremath{2\omega} 
field exhibits more asymmetry in its temporal shape. An increase in the valley polarization as 
compared to the single-color value indicates that the addition of the \ensuremath{2\omega} field significantly affects the valley polarization. 
The phase dependence of the valley polarization shows that the \ensuremath{\varphi}$_{rel}$ between \ensuremath{\omega} 
and \ensuremath{2\omega} is crucial to tune the valley polarization. 
It is also worth mentioning that the \ensuremath{\varphi}$_{rel}$ dependence changes according to the phase of the fundamental pulse. 
The valley 
polarization dependence on \ensuremath{\varphi}$_{rel}$ can be 
explained by the asymmetry of the vector potential and electric field as shown in 
Fig. \ref{fig:4}(b) and (c). Here the asymmetry of the 
vector potential and the electric field is defined as
\begin{equation}
Asym_{A(E)}=\left|\frac{A_{+}(E_{+})(t)}{A_{-}(E_{-})(t)}\right|\label{GrindEQ__10_-1},
\end{equation}
where $A_{\pm}(t)$ and $E_{\pm}(t)$ are the positive ($+$) and negative ($-$) peaks of
the vector potential and electric field respectively. 
%Note that A(t) describes how large area is related to 
%excitation while E(t) depicts the excitation rate.

\begin{figure*}
\centering \includegraphics[scale=0.55]{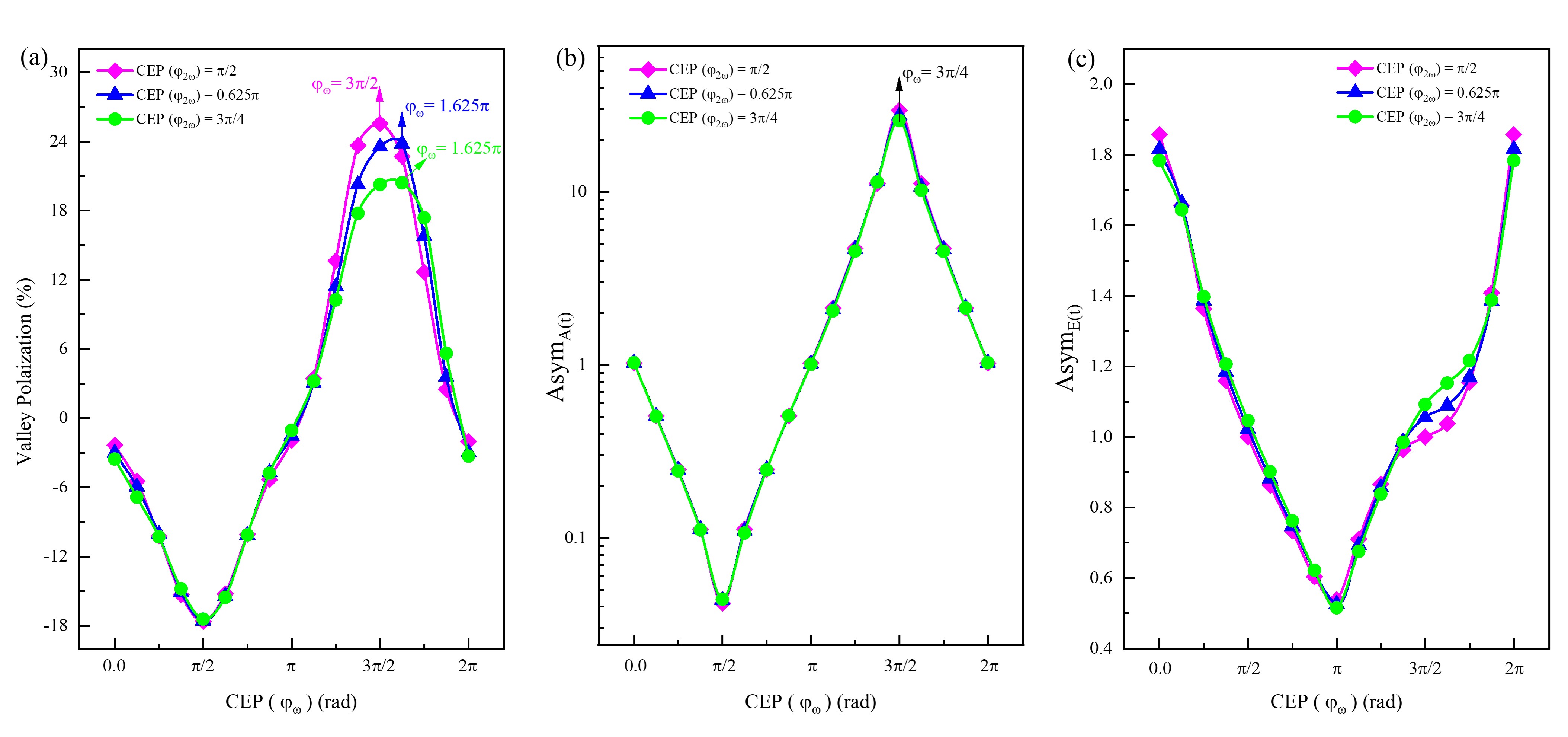} \caption{(a) Valley polarization (b) $Asym_{A(t)}$ and (c) $Asym_{E(t)}$ as a function of \ensuremath{\varphi}$_{\omega}$ at maximum valley polarization of \ensuremath{\varphi}$_{2\omega}$ in Fig. \ref{fig:4}(a). }
\label{fig:5} 
\end{figure*}

Valley polarization and $Asym_{A(t)}$ have strong 
relevance. i.e smaller vector potential value produces 
weaker valley polarization while a higher value of
$Asym_{A(t)}$ results in a stronger valley polarization. 
In our calculation, the $Asym_{A(t)}$ (asymmetry of excitation area) dictates 
the main peak position of valley polarization. However, the peak of the valley 
polarization may not directly correspond to the maxima of the vector potential. 
As one can see that the valley polarization peak for 
\ensuremath{\varphi}$_{\omega}$ = \ensuremath{1.625\pi} is at \ensuremath{0.625\pi} 
while the $Asym_{A(t)}$ peak is at \ensuremath{0.375\pi} of \ensuremath{\varphi}$_{2\omega}$ 
in Fig. \ref{fig:5}. Similarly for \ensuremath{\varphi}$_{\omega}$ = \ensuremath{7\pi}/4 
the valley peak is at \ensuremath{3\pi}/4 while the $Asym_{A(t)}$ peak is 
at \ensuremath{\pi}/2 of \ensuremath{\varphi}$_{2\omega}$.
In principle, the peak of valley polarization should follow the $Asym_{A(t)}$ 
if the tunneling of the electrons is solely responsible for the valley polarization. 
Even though it is a minor effect, the asymmetric excitation
rate due to E(t) influences valley polarization. 
Since the $Asym_{E(t)}$ is shifted by CEP, the asymmetry of valley polarization shifts slightly by changing CEP.

The peak value of valley polarization in Fig. \ref{fig:4}(a) 
is observed at \ensuremath{\varphi}$_{2\omega}$ = \ensuremath{\pi}/2, \ensuremath{0.625\pi} and \ensuremath{3\pi}/4. 
Thus, we have explored the valley polarization at the peaks 
of \ensuremath{\varphi}$_{2\omega}$ as a function 
of \ensuremath{\varphi}$_{\omega}$. 
Fig. \ref{fig:5}(a) shows the valley polarization as a function of 
the \ensuremath{\varphi}$_{\omega}$ at the peaks of \ensuremath{\varphi}$_{2\omega}$. 
The valley polarization curve at various \ensuremath{\varphi}$_{2\omega}$ nearly 
overlap. 
The \ensuremath{\varphi}$_{\omega}$ dependence of 
valley polarization shows the same behavior as in the case of a single-color linearly 
polarized pulse\citep{jimenez2021sub,hashmi2022valley}, except an increase in 
the valley polarization value in the second positive half of the sine-like curve.  
Thus, this valley polarization increase is due to the \ensuremath{2\omega}-field. 
The $Asym_{A(t)}$ in Fig. \ref{fig:5}(b) and 
the $Asym_{E(t)}$ in Fig. \ref{fig:5}(c) show very distinctive curves. Due to the predominant
tunneling nature of \ensuremath{\omega}-field, the valley polarization nearly
follows the $Asym_{A(t)}$ curve as shown in Fig.~\ref{fig:5}(b). The change in the shoulder structure of $Asym_{E(t)}$ in Fig.~\ref{fig:5}(c) 
changes the peak position 
of valley asymmetry to \ensuremath{\varphi}$_{\omega}$ = \ensuremath{1.625\pi} for \ensuremath{\varphi}$_{2\omega}$ = \ensuremath{0.625\pi} and \ensuremath{3\pi}/4 which marks
the importance of $Asym_{E(t)}$ by \ensuremath{\omega} + \ensuremath{2\omega} field. 
Thus the competition between the asymmetry of excitation 
rate and excitation area by
the addition of \ensuremath{2\omega}-field determines the peak of valley polarization.

\section{CONCLUSION}

We have demonstrated the generation of valley polarization in WSe$_{2}$ monolayer via \ensuremath{\omega} + \ensuremath{2\omega} superimposed 
co-linearly laser pulses.  
The valley polarization 
induced by \ensuremath{\omega} + \ensuremath{2\omega} pulses 
is superior to single-color pulse as much as 1.2 times.
The intensity ratio and the relative phase between 
the fundamental pulse and its second harmonic are the most 
crucial factors in determining valley polarization. 
Our results indicate that the strong \ensuremath{\omega} 
mixed with the weaker \ensuremath{2\omega} enhances the valley polarization. 
We found that the ratio of $I_{\omega}/I_{2\omega}=36$ shows higher valley polarization
than the ratio used in molecule cases.
%Interestingly, the ratio of $E_{\omega}/E_{2\omega}$ in WSe$_{2}$ monolayer is found to be 6 
%which is much higher than the vacuum ($E_{\omega}/E_{2\omega}$ = 3). 
The valley polarization is mainly dependent on the CEP of the fundamental pulse. 
The competition between the asymmetry of the vector potential and electric field mainly determined 
the maximum valley polarization. 
Two-color fields provide additional degrees of freedom, such as admixture, polarization, and the relative phase between two pulses. 
The present scheme helps to understand the ultrafast valley phenomena in 2D layers from the fundamental perspective.

\section{AUTHOR INFORMATION}

\noindent \textbf{Corresponding Author}

\noindent {*}E-mail: otobe.tomohito@qst.go.jp

\noindent \textbf{Author Contributions}

%$\mathrm{\dagger}$ A. H. and S.Y contributed equally to this work.
%\noindent T.O. conceived the idea and supervised this study. A.H performed TDDFT calculations, K.Y. developed the massive Dirac Hamiltonian, S.Y. implemented the two band-model and spin-orbit included methods of projection to analyze the excited states. All authors have contributed to the writing and the interpretation of results.\textbf{}

\noindent \textbf{Notes}

\noindent The authors declare no competing financial interest.

\section{ACKNOWLEDGMENT}

\noindent This research is supported by JST-CREST under Grant No.
JP-MJCR16N5. This research is also partially supported by JSPS KAKENHI
Grant Nos. 20H02649, 22K13991, and MEXT Quantum Leap Flagship Program (MEXT Q-LEAP)
under Grant No. JPMXS0118068681 and JPMXS0118067246. The numerical calculations are carried
out using the computer facilities of the Fugaku through the HPCI System
Research Project (Project ID: hp220120), SGI8600 at Japan Atomic Energy
Agency (JAEA), and Multidisciplinary Cooperative Research Program
in CCS, University of Tsukuba.

\bibliography{bibliography}

\begin{thebibliography}{49}%
\makeatletter
\providecommand \@ifxundefined [1]{%
 \@ifx{#1\undefined}
}%
\providecommand \@ifnum [1]{%
 \ifnum #1\expandafter \@firstoftwo
 \else \expandafter \@secondoftwo
 \fi
}%
\providecommand \@ifx [1]{%
 \ifx #1\expandafter \@firstoftwo
 \else \expandafter \@secondoftwo
 \fi
}%
\providecommand \natexlab [1]{#1}%
\providecommand \enquote  [1]{``#1''}%
\providecommand \bibnamefont  [1]{#1}%
\providecommand \bibfnamefont [1]{#1}%
\providecommand \citenamefont [1]{#1}%
\providecommand \href@noop [0]{\@secondoftwo}%
\providecommand \href [0]{\begingroup \@sanitize@url \@href}%
\providecommand \@href[1]{\@@startlink{#1}\@@href}%
\providecommand \@@href[1]{\endgroup#1\@@endlink}%
\providecommand \@sanitize@url [0]{\catcode `\\12\catcode `\$12\catcode
  `\&12\catcode `\#12\catcode `\^12\catcode `\_12\catcode `\%12\relax}%
\providecommand \@@startlink[1]{}%
\providecommand \@@endlink[0]{}%
\providecommand \url  [0]{\begingroup\@sanitize@url \@url }%
\providecommand \@url [1]{\endgroup\@href {#1}{\urlprefix }}%
\providecommand \urlprefix  [0]{URL }%
\providecommand \Eprint [0]{\href }%
\providecommand \doibase [0]{https://doi.org/}%
\providecommand \selectlanguage [0]{\@gobble}%
\providecommand \bibinfo  [0]{\@secondoftwo}%
\providecommand \bibfield  [0]{\@secondoftwo}%
\providecommand \translation [1]{[#1]}%
\providecommand \BibitemOpen [0]{}%
\providecommand \bibitemStop [0]{}%
\providecommand \bibitemNoStop [0]{.\EOS\space}%
\providecommand \EOS [0]{\spacefactor3000\relax}%
\providecommand \BibitemShut  [1]{\csname bibitem#1\endcsname}%
\let\auto@bib@innerbib\@empty
%</preamble>
\bibitem [{\citenamefont {Xiao}\ \emph {et~al.}(2012)\citenamefont {Xiao},
  \citenamefont {Liu}, \citenamefont {Feng}, \citenamefont {Xu},\ and\
  \citenamefont {Yao}}]{xiao2012coupled}%
  \BibitemOpen
  \bibfield  {author} {\bibinfo {author} {\bibfnamefont {D.}~\bibnamefont
  {Xiao}}, \bibinfo {author} {\bibfnamefont {G.-B.}\ \bibnamefont {Liu}},
  \bibinfo {author} {\bibfnamefont {W.}~\bibnamefont {Feng}}, \bibinfo {author}
  {\bibfnamefont {X.}~\bibnamefont {Xu}},\ and\ \bibinfo {author}
  {\bibfnamefont {W.}~\bibnamefont {Yao}},\ }\bibfield  {title} {\bibinfo
  {title} {Coupled spin and valley physics in monolayers of mos 2 and other
  group-vi dichalcogenides},\ }\href@noop {} {\bibfield  {journal} {\bibinfo
  {journal} {Physical review letters}\ }\textbf {\bibinfo {volume} {108}},\
  \bibinfo {pages} {196802} (\bibinfo {year} {2012})}\BibitemShut {NoStop}%
\bibitem [{\citenamefont {Xu}\ \emph {et~al.}(2014)\citenamefont {Xu},
  \citenamefont {Yao}, \citenamefont {Xiao},\ and\ \citenamefont
  {Heinz}}]{xu2014spin}%
  \BibitemOpen
  \bibfield  {author} {\bibinfo {author} {\bibfnamefont {X.}~\bibnamefont
  {Xu}}, \bibinfo {author} {\bibfnamefont {W.}~\bibnamefont {Yao}}, \bibinfo
  {author} {\bibfnamefont {D.}~\bibnamefont {Xiao}},\ and\ \bibinfo {author}
  {\bibfnamefont {T.~F.}\ \bibnamefont {Heinz}},\ }\bibfield  {title} {\bibinfo
  {title} {Spin and pseudospins in layered transition metal dichalcogenides},\
  }\href@noop {} {\bibfield  {journal} {\bibinfo  {journal} {Nature Physics}\
  }\textbf {\bibinfo {volume} {10}},\ \bibinfo {pages} {343} (\bibinfo {year}
  {2014})}\BibitemShut {NoStop}%
\bibitem [{\citenamefont {Wang}\ \emph {et~al.}(2012)\citenamefont {Wang},
  \citenamefont {Kalantar-Zadeh}, \citenamefont {Kis}, \citenamefont
  {Coleman},\ and\ \citenamefont {Strano}}]{wang2012electronics}%
  \BibitemOpen
  \bibfield  {author} {\bibinfo {author} {\bibfnamefont {Q.~H.}\ \bibnamefont
  {Wang}}, \bibinfo {author} {\bibfnamefont {K.}~\bibnamefont
  {Kalantar-Zadeh}}, \bibinfo {author} {\bibfnamefont {A.}~\bibnamefont {Kis}},
  \bibinfo {author} {\bibfnamefont {J.~N.}\ \bibnamefont {Coleman}},\ and\
  \bibinfo {author} {\bibfnamefont {M.~S.}\ \bibnamefont {Strano}},\ }\bibfield
   {title} {\bibinfo {title} {Electronics and optoelectronics of
  two-dimensional transition metal dichalcogenides},\ }\href@noop {} {\bibfield
   {journal} {\bibinfo  {journal} {Nature nanotechnology}\ }\textbf {\bibinfo
  {volume} {7}},\ \bibinfo {pages} {699} (\bibinfo {year} {2012})}\BibitemShut
  {NoStop}%
\bibitem [{\citenamefont {Schaibley}\ \emph {et~al.}(2016)\citenamefont
  {Schaibley}, \citenamefont {Yu}, \citenamefont {Clark}, \citenamefont
  {Rivera}, \citenamefont {Ross}, \citenamefont {Seyler}, \citenamefont {Yao},\
  and\ \citenamefont {Xu}}]{schaibley2016valleytronics}%
  \BibitemOpen
  \bibfield  {author} {\bibinfo {author} {\bibfnamefont {J.~R.}\ \bibnamefont
  {Schaibley}}, \bibinfo {author} {\bibfnamefont {H.}~\bibnamefont {Yu}},
  \bibinfo {author} {\bibfnamefont {G.}~\bibnamefont {Clark}}, \bibinfo
  {author} {\bibfnamefont {P.}~\bibnamefont {Rivera}}, \bibinfo {author}
  {\bibfnamefont {J.~S.}\ \bibnamefont {Ross}}, \bibinfo {author}
  {\bibfnamefont {K.~L.}\ \bibnamefont {Seyler}}, \bibinfo {author}
  {\bibfnamefont {W.}~\bibnamefont {Yao}},\ and\ \bibinfo {author}
  {\bibfnamefont {X.}~\bibnamefont {Xu}},\ }\bibfield  {title} {\bibinfo
  {title} {Valleytronics in 2d materials},\ }\href@noop {} {\bibfield
  {journal} {\bibinfo  {journal} {Nature Reviews Materials}\ }\textbf {\bibinfo
  {volume} {1}},\ \bibinfo {pages} {1} (\bibinfo {year} {2016})}\BibitemShut
  {NoStop}%
\bibitem [{\citenamefont {Xiao}\ \emph {et~al.}(2007)\citenamefont {Xiao},
  \citenamefont {Yao},\ and\ \citenamefont {Niu}}]{xiao2007valley}%
  \BibitemOpen
  \bibfield  {author} {\bibinfo {author} {\bibfnamefont {D.}~\bibnamefont
  {Xiao}}, \bibinfo {author} {\bibfnamefont {W.}~\bibnamefont {Yao}},\ and\
  \bibinfo {author} {\bibfnamefont {Q.}~\bibnamefont {Niu}},\ }\bibfield
  {title} {\bibinfo {title} {Valley-contrasting physics in graphene: magnetic
  moment and topological transport},\ }\href@noop {} {\bibfield  {journal}
  {\bibinfo  {journal} {Physical review letters}\ }\textbf {\bibinfo {volume}
  {99}},\ \bibinfo {pages} {236809} (\bibinfo {year} {2007})}\BibitemShut
  {NoStop}%
\bibitem [{\citenamefont {Shimazaki}\ \emph {et~al.}(2015)\citenamefont
  {Shimazaki}, \citenamefont {Yamamoto}, \citenamefont {Borzenets},
  \citenamefont {Watanabe}, \citenamefont {Taniguchi},\ and\ \citenamefont
  {Tarucha}}]{shimazaki2015generation}%
  \BibitemOpen
  \bibfield  {author} {\bibinfo {author} {\bibfnamefont {Y.}~\bibnamefont
  {Shimazaki}}, \bibinfo {author} {\bibfnamefont {M.}~\bibnamefont {Yamamoto}},
  \bibinfo {author} {\bibfnamefont {I.~V.}\ \bibnamefont {Borzenets}}, \bibinfo
  {author} {\bibfnamefont {K.}~\bibnamefont {Watanabe}}, \bibinfo {author}
  {\bibfnamefont {T.}~\bibnamefont {Taniguchi}},\ and\ \bibinfo {author}
  {\bibfnamefont {S.}~\bibnamefont {Tarucha}},\ }\bibfield  {title} {\bibinfo
  {title} {Generation and detection of pure valley current by electrically
  induced berry curvature in bilayer graphene},\ }\href@noop {} {\bibfield
  {journal} {\bibinfo  {journal} {Nature Physics}\ }\textbf {\bibinfo {volume}
  {11}},\ \bibinfo {pages} {1032} (\bibinfo {year} {2015})}\BibitemShut
  {NoStop}%
\bibitem [{\citenamefont {Sie}(2018)}]{sie2018valley}%
  \BibitemOpen
  \bibfield  {author} {\bibinfo {author} {\bibfnamefont {E.~J.}\ \bibnamefont
  {Sie}},\ }\bibfield  {title} {\bibinfo {title} {Valley-selective optical
  stark effect in monolayer ws 2},\ }in\ \href@noop {} {\emph {\bibinfo
  {booktitle} {Coherent Light-Matter Interactions in Monolayer Transition-Metal
  Dichalcogenides}}}\ (\bibinfo  {publisher} {Springer},\ \bibinfo {year}
  {2018})\ pp.\ \bibinfo {pages} {37--57}\BibitemShut {NoStop}%
\bibitem [{\citenamefont {Cao}\ \emph {et~al.}(2012)\citenamefont {Cao},
  \citenamefont {Wang}, \citenamefont {Han}, \citenamefont {Ye}, \citenamefont
  {Zhu}, \citenamefont {Shi}, \citenamefont {Niu}, \citenamefont {Tan},
  \citenamefont {Wang}, \citenamefont {Liu} \emph {et~al.}}]{cao2012valley}%
  \BibitemOpen
  \bibfield  {author} {\bibinfo {author} {\bibfnamefont {T.}~\bibnamefont
  {Cao}}, \bibinfo {author} {\bibfnamefont {G.}~\bibnamefont {Wang}}, \bibinfo
  {author} {\bibfnamefont {W.}~\bibnamefont {Han}}, \bibinfo {author}
  {\bibfnamefont {H.}~\bibnamefont {Ye}}, \bibinfo {author} {\bibfnamefont
  {C.}~\bibnamefont {Zhu}}, \bibinfo {author} {\bibfnamefont {J.}~\bibnamefont
  {Shi}}, \bibinfo {author} {\bibfnamefont {Q.}~\bibnamefont {Niu}}, \bibinfo
  {author} {\bibfnamefont {P.}~\bibnamefont {Tan}}, \bibinfo {author}
  {\bibfnamefont {E.}~\bibnamefont {Wang}}, \bibinfo {author} {\bibfnamefont
  {B.}~\bibnamefont {Liu}}, \emph {et~al.},\ }\bibfield  {title} {\bibinfo
  {title} {Valley-selective circular dichroism of monolayer molybdenum
  disulphide},\ }\href@noop {} {\bibfield  {journal} {\bibinfo  {journal}
  {Nature communications}\ }\textbf {\bibinfo {volume} {3}},\ \bibinfo {pages}
  {1} (\bibinfo {year} {2012})}\BibitemShut {NoStop}%
\bibitem [{\citenamefont {Jones}\ \emph {et~al.}(2013)\citenamefont {Jones},
  \citenamefont {Yu}, \citenamefont {Ghimire}, \citenamefont {Wu},
  \citenamefont {Aivazian}, \citenamefont {Ross}, \citenamefont {Zhao},
  \citenamefont {Yan}, \citenamefont {Mandrus}, \citenamefont {Xiao} \emph
  {et~al.}}]{jones2013optical}%
  \BibitemOpen
  \bibfield  {author} {\bibinfo {author} {\bibfnamefont {A.~M.}\ \bibnamefont
  {Jones}}, \bibinfo {author} {\bibfnamefont {H.}~\bibnamefont {Yu}}, \bibinfo
  {author} {\bibfnamefont {N.~J.}\ \bibnamefont {Ghimire}}, \bibinfo {author}
  {\bibfnamefont {S.}~\bibnamefont {Wu}}, \bibinfo {author} {\bibfnamefont
  {G.}~\bibnamefont {Aivazian}}, \bibinfo {author} {\bibfnamefont {J.~S.}\
  \bibnamefont {Ross}}, \bibinfo {author} {\bibfnamefont {B.}~\bibnamefont
  {Zhao}}, \bibinfo {author} {\bibfnamefont {J.}~\bibnamefont {Yan}}, \bibinfo
  {author} {\bibfnamefont {D.~G.}\ \bibnamefont {Mandrus}}, \bibinfo {author}
  {\bibfnamefont {D.}~\bibnamefont {Xiao}}, \emph {et~al.},\ }\bibfield
  {title} {\bibinfo {title} {Optical generation of excitonic valley coherence
  in monolayer wse2},\ }\href@noop {} {\bibfield  {journal} {\bibinfo
  {journal} {Nature nanotechnology}\ }\textbf {\bibinfo {volume} {8}},\
  \bibinfo {pages} {634} (\bibinfo {year} {2013})}\BibitemShut {NoStop}%
\bibitem [{\citenamefont {Geim}(2009)}]{geim2009graphene}%
  \BibitemOpen
  \bibfield  {author} {\bibinfo {author} {\bibfnamefont {A.~K.}\ \bibnamefont
  {Geim}},\ }\bibfield  {title} {\bibinfo {title} {Graphene: status and
  prospects},\ }\href@noop {} {\bibfield  {journal} {\bibinfo  {journal}
  {science}\ }\textbf {\bibinfo {volume} {324}},\ \bibinfo {pages} {1530}
  (\bibinfo {year} {2009})}\BibitemShut {NoStop}%
\bibitem [{\citenamefont {Neto}\ \emph {et~al.}(2009)\citenamefont {Neto},
  \citenamefont {Guinea}, \citenamefont {Peres}, \citenamefont {Novoselov},\
  and\ \citenamefont {Geim}}]{neto2009electronic}%
  \BibitemOpen
  \bibfield  {author} {\bibinfo {author} {\bibfnamefont {A.~C.}\ \bibnamefont
  {Neto}}, \bibinfo {author} {\bibfnamefont {F.}~\bibnamefont {Guinea}},
  \bibinfo {author} {\bibfnamefont {N.~M.}\ \bibnamefont {Peres}}, \bibinfo
  {author} {\bibfnamefont {K.~S.}\ \bibnamefont {Novoselov}},\ and\ \bibinfo
  {author} {\bibfnamefont {A.~K.}\ \bibnamefont {Geim}},\ }\bibfield  {title}
  {\bibinfo {title} {The electronic properties of graphene},\ }\href@noop {}
  {\bibfield  {journal} {\bibinfo  {journal} {Reviews of modern physics}\
  }\textbf {\bibinfo {volume} {81}},\ \bibinfo {pages} {109} (\bibinfo {year}
  {2009})}\BibitemShut {NoStop}%
\bibitem [{\citenamefont {Zhang}\ \emph {et~al.}(2005)\citenamefont {Zhang},
  \citenamefont {Tan}, \citenamefont {Stormer},\ and\ \citenamefont
  {Kim}}]{zhang2005experimental}%
  \BibitemOpen
  \bibfield  {author} {\bibinfo {author} {\bibfnamefont {Y.}~\bibnamefont
  {Zhang}}, \bibinfo {author} {\bibfnamefont {Y.-W.}\ \bibnamefont {Tan}},
  \bibinfo {author} {\bibfnamefont {H.~L.}\ \bibnamefont {Stormer}},\ and\
  \bibinfo {author} {\bibfnamefont {P.}~\bibnamefont {Kim}},\ }\bibfield
  {title} {\bibinfo {title} {Experimental observation of the quantum hall
  effect and berry's phase in graphene},\ }\href@noop {} {\bibfield  {journal}
  {\bibinfo  {journal} {nature}\ }\textbf {\bibinfo {volume} {438}},\ \bibinfo
  {pages} {201} (\bibinfo {year} {2005})}\BibitemShut {NoStop}%
\bibitem [{\citenamefont {Yuan}\ \emph {et~al.}(2014)\citenamefont {Yuan},
  \citenamefont {Wang}, \citenamefont {Lian}, \citenamefont {Zhang},
  \citenamefont {Fang}, \citenamefont {Shen}, \citenamefont {Xu}, \citenamefont
  {Xu}, \citenamefont {Zhang}, \citenamefont {Hwang} \emph
  {et~al.}}]{yuan2014generation}%
  \BibitemOpen
  \bibfield  {author} {\bibinfo {author} {\bibfnamefont {H.}~\bibnamefont
  {Yuan}}, \bibinfo {author} {\bibfnamefont {X.}~\bibnamefont {Wang}}, \bibinfo
  {author} {\bibfnamefont {B.}~\bibnamefont {Lian}}, \bibinfo {author}
  {\bibfnamefont {H.}~\bibnamefont {Zhang}}, \bibinfo {author} {\bibfnamefont
  {X.}~\bibnamefont {Fang}}, \bibinfo {author} {\bibfnamefont {B.}~\bibnamefont
  {Shen}}, \bibinfo {author} {\bibfnamefont {G.}~\bibnamefont {Xu}}, \bibinfo
  {author} {\bibfnamefont {Y.}~\bibnamefont {Xu}}, \bibinfo {author}
  {\bibfnamefont {S.-C.}\ \bibnamefont {Zhang}}, \bibinfo {author}
  {\bibfnamefont {H.~Y.}\ \bibnamefont {Hwang}}, \emph {et~al.},\ }\bibfield
  {title} {\bibinfo {title} {Generation and electric control of
  spin--valley-coupled circular photogalvanic current in wse2},\ }\href@noop {}
  {\bibfield  {journal} {\bibinfo  {journal} {Nature nanotechnology}\ }\textbf
  {\bibinfo {volume} {9}},\ \bibinfo {pages} {851} (\bibinfo {year}
  {2014})}\BibitemShut {NoStop}%
\bibitem [{\citenamefont {Vitale}\ \emph {et~al.}(2018)\citenamefont {Vitale},
  \citenamefont {Nezich}, \citenamefont {Varghese}, \citenamefont {Kim},
  \citenamefont {Gedik}, \citenamefont {Jarillo-Herrero}, \citenamefont
  {Xiao},\ and\ \citenamefont {Rothschild}}]{vitale2018valleytronics}%
  \BibitemOpen
  \bibfield  {author} {\bibinfo {author} {\bibfnamefont {S.~A.}\ \bibnamefont
  {Vitale}}, \bibinfo {author} {\bibfnamefont {D.}~\bibnamefont {Nezich}},
  \bibinfo {author} {\bibfnamefont {J.~O.}\ \bibnamefont {Varghese}}, \bibinfo
  {author} {\bibfnamefont {P.}~\bibnamefont {Kim}}, \bibinfo {author}
  {\bibfnamefont {N.}~\bibnamefont {Gedik}}, \bibinfo {author} {\bibfnamefont
  {P.}~\bibnamefont {Jarillo-Herrero}}, \bibinfo {author} {\bibfnamefont
  {D.}~\bibnamefont {Xiao}},\ and\ \bibinfo {author} {\bibfnamefont
  {M.}~\bibnamefont {Rothschild}},\ }\bibfield  {title} {\bibinfo {title}
  {Valleytronics: opportunities, challenges, and paths forward},\ }\href@noop
  {} {\bibfield  {journal} {\bibinfo  {journal} {Small}\ }\textbf {\bibinfo
  {volume} {14}},\ \bibinfo {pages} {1801483} (\bibinfo {year}
  {2018})}\BibitemShut {NoStop}%
\bibitem [{\citenamefont {MacNeill}\ \emph {et~al.}(2015)\citenamefont
  {MacNeill}, \citenamefont {Heikes}, \citenamefont {Mak}, \citenamefont
  {Anderson}, \citenamefont {Korm{\'a}nyos}, \citenamefont {Z{\'o}lyomi},
  \citenamefont {Park},\ and\ \citenamefont {Ralph}}]{macneill2015breaking}%
  \BibitemOpen
  \bibfield  {author} {\bibinfo {author} {\bibfnamefont {D.}~\bibnamefont
  {MacNeill}}, \bibinfo {author} {\bibfnamefont {C.}~\bibnamefont {Heikes}},
  \bibinfo {author} {\bibfnamefont {K.~F.}\ \bibnamefont {Mak}}, \bibinfo
  {author} {\bibfnamefont {Z.}~\bibnamefont {Anderson}}, \bibinfo {author}
  {\bibfnamefont {A.}~\bibnamefont {Korm{\'a}nyos}}, \bibinfo {author}
  {\bibfnamefont {V.}~\bibnamefont {Z{\'o}lyomi}}, \bibinfo {author}
  {\bibfnamefont {J.}~\bibnamefont {Park}},\ and\ \bibinfo {author}
  {\bibfnamefont {D.~C.}\ \bibnamefont {Ralph}},\ }\bibfield  {title} {\bibinfo
  {title} {Breaking of valley degeneracy by magnetic field in monolayer mose
  2},\ }\href@noop {} {\bibfield  {journal} {\bibinfo  {journal} {Physical
  review letters}\ }\textbf {\bibinfo {volume} {114}},\ \bibinfo {pages}
  {037401} (\bibinfo {year} {2015})}\BibitemShut {NoStop}%
\bibitem [{\citenamefont {Kim}\ \emph {et~al.}(2014)\citenamefont {Kim},
  \citenamefont {Hong}, \citenamefont {Jin}, \citenamefont {Shi}, \citenamefont
  {Chang}, \citenamefont {Chiu}, \citenamefont {Li},\ and\ \citenamefont
  {Wang}}]{kim2014ultrafast}%
  \BibitemOpen
  \bibfield  {author} {\bibinfo {author} {\bibfnamefont {J.}~\bibnamefont
  {Kim}}, \bibinfo {author} {\bibfnamefont {X.}~\bibnamefont {Hong}}, \bibinfo
  {author} {\bibfnamefont {C.}~\bibnamefont {Jin}}, \bibinfo {author}
  {\bibfnamefont {S.-F.}\ \bibnamefont {Shi}}, \bibinfo {author} {\bibfnamefont
  {C.-Y.~S.}\ \bibnamefont {Chang}}, \bibinfo {author} {\bibfnamefont {M.-H.}\
  \bibnamefont {Chiu}}, \bibinfo {author} {\bibfnamefont {L.-J.}\ \bibnamefont
  {Li}},\ and\ \bibinfo {author} {\bibfnamefont {F.}~\bibnamefont {Wang}},\
  }\bibfield  {title} {\bibinfo {title} {Ultrafast generation of
  pseudo-magnetic field for valley excitons in wse2 monolayers},\ }\href@noop
  {} {\bibfield  {journal} {\bibinfo  {journal} {Science}\ }\textbf {\bibinfo
  {volume} {346}},\ \bibinfo {pages} {1205} (\bibinfo {year}
  {2014})}\BibitemShut {NoStop}%
\bibitem [{\citenamefont {Zhong}\ \emph {et~al.}(2017)\citenamefont {Zhong},
  \citenamefont {Seyler}, \citenamefont {Linpeng}, \citenamefont {Cheng},
  \citenamefont {Sivadas}, \citenamefont {Huang}, \citenamefont {Schmidgall},
  \citenamefont {Taniguchi}, \citenamefont {Watanabe}, \citenamefont {McGuire}
  \emph {et~al.}}]{zhong2017van}%
  \BibitemOpen
  \bibfield  {author} {\bibinfo {author} {\bibfnamefont {D.}~\bibnamefont
  {Zhong}}, \bibinfo {author} {\bibfnamefont {K.~L.}\ \bibnamefont {Seyler}},
  \bibinfo {author} {\bibfnamefont {X.}~\bibnamefont {Linpeng}}, \bibinfo
  {author} {\bibfnamefont {R.}~\bibnamefont {Cheng}}, \bibinfo {author}
  {\bibfnamefont {N.}~\bibnamefont {Sivadas}}, \bibinfo {author} {\bibfnamefont
  {B.}~\bibnamefont {Huang}}, \bibinfo {author} {\bibfnamefont
  {E.}~\bibnamefont {Schmidgall}}, \bibinfo {author} {\bibfnamefont
  {T.}~\bibnamefont {Taniguchi}}, \bibinfo {author} {\bibfnamefont
  {K.}~\bibnamefont {Watanabe}}, \bibinfo {author} {\bibfnamefont {M.~A.}\
  \bibnamefont {McGuire}}, \emph {et~al.},\ }\bibfield  {title} {\bibinfo
  {title} {Van der waals engineering of ferromagnetic semiconductor
  heterostructures for spin and valleytronics},\ }\href@noop {} {\bibfield
  {journal} {\bibinfo  {journal} {Science advances}\ }\textbf {\bibinfo
  {volume} {3}},\ \bibinfo {pages} {e1603113} (\bibinfo {year}
  {2017})}\BibitemShut {NoStop}%
\bibitem [{\citenamefont {Zeng}\ \emph {et~al.}(2012)\citenamefont {Zeng},
  \citenamefont {Dai}, \citenamefont {Yao}, \citenamefont {Xiao},\ and\
  \citenamefont {Cui}}]{zeng2012valley}%
  \BibitemOpen
  \bibfield  {author} {\bibinfo {author} {\bibfnamefont {H.}~\bibnamefont
  {Zeng}}, \bibinfo {author} {\bibfnamefont {J.}~\bibnamefont {Dai}}, \bibinfo
  {author} {\bibfnamefont {W.}~\bibnamefont {Yao}}, \bibinfo {author}
  {\bibfnamefont {D.}~\bibnamefont {Xiao}},\ and\ \bibinfo {author}
  {\bibfnamefont {X.}~\bibnamefont {Cui}},\ }\bibfield  {title} {\bibinfo
  {title} {Valley polarization in mos 2 monolayers by optical pumping},\
  }\href@noop {} {\bibfield  {journal} {\bibinfo  {journal} {Nature
  nanotechnology}\ }\textbf {\bibinfo {volume} {7}},\ \bibinfo {pages} {490}
  (\bibinfo {year} {2012})}\BibitemShut {NoStop}%
\bibitem [{\citenamefont {Mak}\ \emph {et~al.}(2012)\citenamefont {Mak},
  \citenamefont {He}, \citenamefont {Shan},\ and\ \citenamefont
  {Heinz}}]{mak2012control}%
  \BibitemOpen
  \bibfield  {author} {\bibinfo {author} {\bibfnamefont {K.~F.}\ \bibnamefont
  {Mak}}, \bibinfo {author} {\bibfnamefont {K.}~\bibnamefont {He}}, \bibinfo
  {author} {\bibfnamefont {J.}~\bibnamefont {Shan}},\ and\ \bibinfo {author}
  {\bibfnamefont {T.~F.}\ \bibnamefont {Heinz}},\ }\bibfield  {title} {\bibinfo
  {title} {Control of valley polarization in monolayer mos 2 by optical
  helicity},\ }\href@noop {} {\bibfield  {journal} {\bibinfo  {journal} {Nature
  nanotechnology}\ }\textbf {\bibinfo {volume} {7}},\ \bibinfo {pages} {494}
  (\bibinfo {year} {2012})}\BibitemShut {NoStop}%
\bibitem [{\citenamefont {Yao}\ \emph {et~al.}(2008)\citenamefont {Yao},
  \citenamefont {Xiao},\ and\ \citenamefont {Niu}}]{yao2008valley}%
  \BibitemOpen
  \bibfield  {author} {\bibinfo {author} {\bibfnamefont {W.}~\bibnamefont
  {Yao}}, \bibinfo {author} {\bibfnamefont {D.}~\bibnamefont {Xiao}},\ and\
  \bibinfo {author} {\bibfnamefont {Q.}~\bibnamefont {Niu}},\ }\bibfield
  {title} {\bibinfo {title} {Valley-dependent optoelectronics from inversion
  symmetry breaking},\ }\href@noop {} {\bibfield  {journal} {\bibinfo
  {journal} {Physical Review B}\ }\textbf {\bibinfo {volume} {77}},\ \bibinfo
  {pages} {235406} (\bibinfo {year} {2008})}\BibitemShut {NoStop}%
\bibitem [{\citenamefont {Jim{\'e}nez-Gal{\'a}n}\ \emph
  {et~al.}(2021)\citenamefont {Jim{\'e}nez-Gal{\'a}n}, \citenamefont {Silva},
  \citenamefont {Smirnova},\ and\ \citenamefont {Ivanov}}]{jimenez2021sub}%
  \BibitemOpen
  \bibfield  {author} {\bibinfo {author} {\bibfnamefont {{\'A}.}~\bibnamefont
  {Jim{\'e}nez-Gal{\'a}n}}, \bibinfo {author} {\bibfnamefont {R.~E.}\
  \bibnamefont {Silva}}, \bibinfo {author} {\bibfnamefont {O.}~\bibnamefont
  {Smirnova}},\ and\ \bibinfo {author} {\bibfnamefont {M.}~\bibnamefont
  {Ivanov}},\ }\bibfield  {title} {\bibinfo {title} {Sub-cycle valleytronics:
  control of valley polarization using few-cycle linearly polarized pulses},\
  }\href@noop {} {\bibfield  {journal} {\bibinfo  {journal} {Optica}\ }\textbf
  {\bibinfo {volume} {8}},\ \bibinfo {pages} {277} (\bibinfo {year}
  {2021})}\BibitemShut {NoStop}%
\bibitem [{\citenamefont {Hashmi}\ \emph {et~al.}(2022)\citenamefont {Hashmi},
  \citenamefont {Yamada}, \citenamefont {Yamada}, \citenamefont {Yabana},\ and\
  \citenamefont {Otobe}}]{hashmi2022valley}%
  \BibitemOpen
  \bibfield  {author} {\bibinfo {author} {\bibfnamefont {A.}~\bibnamefont
  {Hashmi}}, \bibinfo {author} {\bibfnamefont {S.}~\bibnamefont {Yamada}},
  \bibinfo {author} {\bibfnamefont {A.}~\bibnamefont {Yamada}}, \bibinfo
  {author} {\bibfnamefont {K.}~\bibnamefont {Yabana}},\ and\ \bibinfo {author}
  {\bibfnamefont {T.}~\bibnamefont {Otobe}},\ }\bibfield  {title} {\bibinfo
  {title} {Valley polarization control in wse 2 monolayer by a single-cycle
  laser pulse},\ }\href@noop {} {\bibfield  {journal} {\bibinfo  {journal}
  {Physical Review B}\ }\textbf {\bibinfo {volume} {105}},\ \bibinfo {pages}
  {115403} (\bibinfo {year} {2022})}\BibitemShut {NoStop}%
\bibitem [{\citenamefont {Langer}\ \emph {et~al.}(2018)\citenamefont {Langer},
  \citenamefont {Schmid}, \citenamefont {Schlauderer}, \citenamefont {Gmitra},
  \citenamefont {Fabian}, \citenamefont {Nagler}, \citenamefont {Sch{\"u}ller},
  \citenamefont {Korn}, \citenamefont {Hawkins}, \citenamefont {Steiner} \emph
  {et~al.}}]{langer2018lightwave}%
  \BibitemOpen
  \bibfield  {author} {\bibinfo {author} {\bibfnamefont {F.}~\bibnamefont
  {Langer}}, \bibinfo {author} {\bibfnamefont {C.~P.}\ \bibnamefont {Schmid}},
  \bibinfo {author} {\bibfnamefont {S.}~\bibnamefont {Schlauderer}}, \bibinfo
  {author} {\bibfnamefont {M.}~\bibnamefont {Gmitra}}, \bibinfo {author}
  {\bibfnamefont {J.}~\bibnamefont {Fabian}}, \bibinfo {author} {\bibfnamefont
  {P.}~\bibnamefont {Nagler}}, \bibinfo {author} {\bibfnamefont
  {C.}~\bibnamefont {Sch{\"u}ller}}, \bibinfo {author} {\bibfnamefont
  {T.}~\bibnamefont {Korn}}, \bibinfo {author} {\bibfnamefont {P.}~\bibnamefont
  {Hawkins}}, \bibinfo {author} {\bibfnamefont {J.}~\bibnamefont {Steiner}},
  \emph {et~al.},\ }\bibfield  {title} {\bibinfo {title} {Lightwave
  valleytronics in a monolayer of tungsten diselenide},\ }\href@noop {}
  {\bibfield  {journal} {\bibinfo  {journal} {Nature}\ }\textbf {\bibinfo
  {volume} {557}},\ \bibinfo {pages} {76} (\bibinfo {year} {2018})}\BibitemShut
  {NoStop}%
\bibitem [{\citenamefont {Kumar}\ \emph {et~al.}(2021)\citenamefont {Kumar},
  \citenamefont {Herath},\ and\ \citenamefont {Apalkov}}]{kumar2021ultrafast}%
  \BibitemOpen
  \bibfield  {author} {\bibinfo {author} {\bibfnamefont {P.}~\bibnamefont
  {Kumar}}, \bibinfo {author} {\bibfnamefont {T.~M.}\ \bibnamefont {Herath}},\
  and\ \bibinfo {author} {\bibfnamefont {V.}~\bibnamefont {Apalkov}},\
  }\bibfield  {title} {\bibinfo {title} {Ultrafast valley polarization in
  bilayer graphene},\ }\href@noop {} {\bibfield  {journal} {\bibinfo  {journal}
  {Journal of Applied Physics}\ }\textbf {\bibinfo {volume} {130}},\ \bibinfo
  {pages} {164301} (\bibinfo {year} {2021})}\BibitemShut {NoStop}%
\bibitem [{\citenamefont {Kelardeh}\ \emph {et~al.}(2022)\citenamefont
  {Kelardeh}, \citenamefont {Saalmann},\ and\ \citenamefont
  {Rost}}]{kelardeh2022ultrashort}%
  \BibitemOpen
  \bibfield  {author} {\bibinfo {author} {\bibfnamefont {H.~K.}\ \bibnamefont
  {Kelardeh}}, \bibinfo {author} {\bibfnamefont {U.}~\bibnamefont {Saalmann}},\
  and\ \bibinfo {author} {\bibfnamefont {J.~M.}\ \bibnamefont {Rost}},\
  }\bibfield  {title} {\bibinfo {title} {Ultrashort laser-driven dynamics of
  massless dirac electrons generating valley polarization in graphene},\
  }\href@noop {} {\bibfield  {journal} {\bibinfo  {journal} {Physical Review
  Research}\ }\textbf {\bibinfo {volume} {4}},\ \bibinfo {pages} {L022014}
  (\bibinfo {year} {2022})}\BibitemShut {NoStop}%
\bibitem [{\citenamefont {Theilhaber}(1992)}]{theilhaber1992ab}%
  \BibitemOpen
  \bibfield  {author} {\bibinfo {author} {\bibfnamefont {J.}~\bibnamefont
  {Theilhaber}},\ }\bibfield  {title} {\bibinfo {title} {Ab initio simulations
  of sodium using time-dependent density-functional theory},\ }\href@noop {}
  {\bibfield  {journal} {\bibinfo  {journal} {Physical Review B}\ }\textbf
  {\bibinfo {volume} {46}},\ \bibinfo {pages} {12990} (\bibinfo {year}
  {1992})}\BibitemShut {NoStop}%
\bibitem [{\citenamefont {Yabana}\ and\ \citenamefont
  {Bertsch}(1996)}]{yabana1996time}%
  \BibitemOpen
  \bibfield  {author} {\bibinfo {author} {\bibfnamefont {K.}~\bibnamefont
  {Yabana}}\ and\ \bibinfo {author} {\bibfnamefont {G.}~\bibnamefont
  {Bertsch}},\ }\bibfield  {title} {\bibinfo {title} {Time-dependent
  local-density approximation in real time},\ }\href@noop {} {\bibfield
  {journal} {\bibinfo  {journal} {Physical Review B}\ }\textbf {\bibinfo
  {volume} {54}},\ \bibinfo {pages} {4484} (\bibinfo {year}
  {1996})}\BibitemShut {NoStop}%
\bibitem [{\citenamefont {Yabana}\ \emph {et~al.}(2006)\citenamefont {Yabana},
  \citenamefont {Nakatsukasa}, \citenamefont {Iwata},\ and\ \citenamefont
  {Bertsch}}]{yabana2006real}%
  \BibitemOpen
  \bibfield  {author} {\bibinfo {author} {\bibfnamefont {K.}~\bibnamefont
  {Yabana}}, \bibinfo {author} {\bibfnamefont {T.}~\bibnamefont {Nakatsukasa}},
  \bibinfo {author} {\bibfnamefont {J.-I.}\ \bibnamefont {Iwata}},\ and\
  \bibinfo {author} {\bibfnamefont {G.}~\bibnamefont {Bertsch}},\ }\bibfield
  {title} {\bibinfo {title} {Real-time, real-space implementation of the linear
  response time-dependent density-functional theory},\ }\href@noop {}
  {\bibfield  {journal} {\bibinfo  {journal} {physica status solidi (b)}\
  }\textbf {\bibinfo {volume} {243}},\ \bibinfo {pages} {1121} (\bibinfo {year}
  {2006})}\BibitemShut {NoStop}%
\bibitem [{\citenamefont {Laurent}\ and\ \citenamefont
  {Jacquemin}(2013)}]{laurent2013td}%
  \BibitemOpen
  \bibfield  {author} {\bibinfo {author} {\bibfnamefont {A.~D.}\ \bibnamefont
  {Laurent}}\ and\ \bibinfo {author} {\bibfnamefont {D.}~\bibnamefont
  {Jacquemin}},\ }\bibfield  {title} {\bibinfo {title} {Td-dft benchmarks: a
  review},\ }\href@noop {} {\bibfield  {journal} {\bibinfo  {journal}
  {International Journal of Quantum Chemistry}\ }\textbf {\bibinfo {volume}
  {113}},\ \bibinfo {pages} {2019} (\bibinfo {year} {2013})}\BibitemShut
  {NoStop}%
\bibitem [{\citenamefont {Castro}\ \emph {et~al.}(2004)\citenamefont {Castro},
  \citenamefont {Marques}, \citenamefont {Alonso}, \citenamefont {Bertsch},\
  and\ \citenamefont {Rubio}}]{castro2004excited}%
  \BibitemOpen
  \bibfield  {author} {\bibinfo {author} {\bibfnamefont {A.}~\bibnamefont
  {Castro}}, \bibinfo {author} {\bibfnamefont {M.~A.}\ \bibnamefont {Marques}},
  \bibinfo {author} {\bibfnamefont {J.~A.}\ \bibnamefont {Alonso}}, \bibinfo
  {author} {\bibfnamefont {G.~F.}\ \bibnamefont {Bertsch}},\ and\ \bibinfo
  {author} {\bibfnamefont {A.}~\bibnamefont {Rubio}},\ }\bibfield  {title}
  {\bibinfo {title} {Excited states dynamics in time-dependent density
  functional theory},\ }\href@noop {} {\bibfield  {journal} {\bibinfo
  {journal} {The European Physical Journal D-Atomic, Molecular, Optical and
  Plasma Physics}\ }\textbf {\bibinfo {volume} {28}},\ \bibinfo {pages} {211}
  (\bibinfo {year} {2004})}\BibitemShut {NoStop}%
\bibitem [{\citenamefont {Nobusada}\ and\ \citenamefont
  {Yabana}(2004)}]{nobusada2004high}%
  \BibitemOpen
  \bibfield  {author} {\bibinfo {author} {\bibfnamefont {K.}~\bibnamefont
  {Nobusada}}\ and\ \bibinfo {author} {\bibfnamefont {K.}~\bibnamefont
  {Yabana}},\ }\bibfield  {title} {\bibinfo {title} {High-order harmonic
  generation from silver clusters: Laser-frequency dependence and the screening
  effect of d electrons},\ }\href@noop {} {\bibfield  {journal} {\bibinfo
  {journal} {Physical Review A}\ }\textbf {\bibinfo {volume} {70}},\ \bibinfo
  {pages} {043411} (\bibinfo {year} {2004})}\BibitemShut {NoStop}%
\bibitem [{\citenamefont {Otobe}\ \emph {et~al.}(2008)\citenamefont {Otobe},
  \citenamefont {Yamagiwa}, \citenamefont {Iwata}, \citenamefont {Yabana},
  \citenamefont {Nakatsukasa},\ and\ \citenamefont {Bertsch}}]{otobe2008first}%
  \BibitemOpen
  \bibfield  {author} {\bibinfo {author} {\bibfnamefont {T.}~\bibnamefont
  {Otobe}}, \bibinfo {author} {\bibfnamefont {M.}~\bibnamefont {Yamagiwa}},
  \bibinfo {author} {\bibfnamefont {J.-I.}\ \bibnamefont {Iwata}}, \bibinfo
  {author} {\bibfnamefont {K.}~\bibnamefont {Yabana}}, \bibinfo {author}
  {\bibfnamefont {T.}~\bibnamefont {Nakatsukasa}},\ and\ \bibinfo {author}
  {\bibfnamefont {G.}~\bibnamefont {Bertsch}},\ }\bibfield  {title} {\bibinfo
  {title} {First-principles electron dynamics simulation for optical breakdown
  of dielectrics under an intense laser field},\ }\href@noop {} {\bibfield
  {journal} {\bibinfo  {journal} {Physical Review B}\ }\textbf {\bibinfo
  {volume} {77}},\ \bibinfo {pages} {165104} (\bibinfo {year}
  {2008})}\BibitemShut {NoStop}%
\bibitem [{\citenamefont {{\v{Z}}uti{\'c}}\ \emph {et~al.}(2004)\citenamefont
  {{\v{Z}}uti{\'c}}, \citenamefont {Fabian},\ and\ \citenamefont
  {Sarma}}]{vzutic2004spintronics}%
  \BibitemOpen
  \bibfield  {author} {\bibinfo {author} {\bibfnamefont {I.}~\bibnamefont
  {{\v{Z}}uti{\'c}}}, \bibinfo {author} {\bibfnamefont {J.}~\bibnamefont
  {Fabian}},\ and\ \bibinfo {author} {\bibfnamefont {S.~D.}\ \bibnamefont
  {Sarma}},\ }\bibfield  {title} {\bibinfo {title} {Spintronics: Fundamentals
  and applications},\ }\href@noop {} {\bibfield  {journal} {\bibinfo  {journal}
  {Reviews of modern physics}\ }\textbf {\bibinfo {volume} {76}},\ \bibinfo
  {pages} {323} (\bibinfo {year} {2004})}\BibitemShut {NoStop}%
\bibitem [{\citenamefont {Ohmura}\ \emph {et~al.}(2014)\citenamefont {Ohmura},
  \citenamefont {Saito},\ and\ \citenamefont {Morishita}}]{Ohmura2014}%
  \BibitemOpen
  \bibfield  {author} {\bibinfo {author} {\bibfnamefont {H.}~\bibnamefont
  {Ohmura}}, \bibinfo {author} {\bibfnamefont {N.}~\bibnamefont {Saito}},\ and\
  \bibinfo {author} {\bibfnamefont {T.}~\bibnamefont {Morishita}},\ }\bibfield
  {title} {\bibinfo {title} {Molecular tunneling ionization of the carbonyl
  sulfide molecule by double-frequency phase-controlled laser fields},\ }\href
  {https://doi.org/10.1103/PhysRevA.89.013405} {\bibfield  {journal} {\bibinfo
  {journal} {Phys. Rev. A}\ }\textbf {\bibinfo {volume} {89}},\ \bibinfo
  {pages} {013405} (\bibinfo {year} {2014})}\BibitemShut {NoStop}%
\bibitem [{\citenamefont {Kaziannis}\ \emph {et~al.}(2014)\citenamefont
  {Kaziannis}, \citenamefont {Kotsina},\ and\ \citenamefont
  {Kosmidis}}]{kaziannis2014interaction}%
  \BibitemOpen
  \bibfield  {author} {\bibinfo {author} {\bibfnamefont {S.}~\bibnamefont
  {Kaziannis}}, \bibinfo {author} {\bibfnamefont {N.}~\bibnamefont {Kotsina}},\
  and\ \bibinfo {author} {\bibfnamefont {C.}~\bibnamefont {Kosmidis}},\
  }\bibfield  {title} {\bibinfo {title} {Interaction of toluene with two-color
  asymmetric laser fields: Controlling the directional emission of molecular
  hydrogen fragments},\ }\href@noop {} {\bibfield  {journal} {\bibinfo
  {journal} {The Journal of Chemical Physics}\ }\textbf {\bibinfo {volume}
  {141}},\ \bibinfo {pages} {104319} (\bibinfo {year} {2014})}\BibitemShut
  {NoStop}%
\bibitem [{\citenamefont {Higuchi}\ \emph {et~al.}(2017)\citenamefont
  {Higuchi}, \citenamefont {Heide}, \citenamefont {Ullmann}, \citenamefont
  {Weber},\ and\ \citenamefont {Hommelhoff}}]{higuchi2017light}%
  \BibitemOpen
  \bibfield  {author} {\bibinfo {author} {\bibfnamefont {T.}~\bibnamefont
  {Higuchi}}, \bibinfo {author} {\bibfnamefont {C.}~\bibnamefont {Heide}},
  \bibinfo {author} {\bibfnamefont {K.}~\bibnamefont {Ullmann}}, \bibinfo
  {author} {\bibfnamefont {H.~B.}\ \bibnamefont {Weber}},\ and\ \bibinfo
  {author} {\bibfnamefont {P.}~\bibnamefont {Hommelhoff}},\ }\bibfield  {title}
  {\bibinfo {title} {Light-field-driven currents in graphene},\ }\href@noop {}
  {\bibfield  {journal} {\bibinfo  {journal} {Nature}\ }\textbf {\bibinfo
  {volume} {550}},\ \bibinfo {pages} {224} (\bibinfo {year}
  {2017})}\BibitemShut {NoStop}%
\bibitem [{\citenamefont {Jim{\'e}nez-Gal{\'a}n}\ \emph
  {et~al.}(2020)\citenamefont {Jim{\'e}nez-Gal{\'a}n}, \citenamefont {Silva},
  \citenamefont {Smirnova},\ and\ \citenamefont
  {Ivanov}}]{jimenez2020lightwave}%
  \BibitemOpen
  \bibfield  {author} {\bibinfo {author} {\bibfnamefont {{\'A}.}~\bibnamefont
  {Jim{\'e}nez-Gal{\'a}n}}, \bibinfo {author} {\bibfnamefont {R.}~\bibnamefont
  {Silva}}, \bibinfo {author} {\bibfnamefont {O.}~\bibnamefont {Smirnova}},\
  and\ \bibinfo {author} {\bibfnamefont {M.}~\bibnamefont {Ivanov}},\
  }\bibfield  {title} {\bibinfo {title} {Lightwave control of topological
  properties in 2d materials for sub-cycle and non-resonant valley
  manipulation},\ }\href@noop {} {\bibfield  {journal} {\bibinfo  {journal}
  {Nature Photonics}\ }\textbf {\bibinfo {volume} {14}},\ \bibinfo {pages}
  {728} (\bibinfo {year} {2020})}\BibitemShut {NoStop}%
\bibitem [{\citenamefont {Bertsch}\ \emph {et~al.}(2000)\citenamefont
  {Bertsch}, \citenamefont {Iwata}, \citenamefont {Rubio},\ and\ \citenamefont
  {Yabana}}]{bertsch2000real}%
  \BibitemOpen
  \bibfield  {author} {\bibinfo {author} {\bibfnamefont {G.~F.}\ \bibnamefont
  {Bertsch}}, \bibinfo {author} {\bibfnamefont {J.-I.}\ \bibnamefont {Iwata}},
  \bibinfo {author} {\bibfnamefont {A.}~\bibnamefont {Rubio}},\ and\ \bibinfo
  {author} {\bibfnamefont {K.}~\bibnamefont {Yabana}},\ }\bibfield  {title}
  {\bibinfo {title} {Real-space, real-time method for the dielectric
  function},\ }\href@noop {} {\bibfield  {journal} {\bibinfo  {journal}
  {Physical Review B}\ }\textbf {\bibinfo {volume} {62}},\ \bibinfo {pages}
  {7998} (\bibinfo {year} {2000})}\BibitemShut {NoStop}%
\bibitem [{\citenamefont {Yamada}\ \emph {et~al.}(2018)\citenamefont {Yamada},
  \citenamefont {Noda}, \citenamefont {Nobusada},\ and\ \citenamefont
  {Yabana}}]{yamada2018time}%
  \BibitemOpen
  \bibfield  {author} {\bibinfo {author} {\bibfnamefont {S.}~\bibnamefont
  {Yamada}}, \bibinfo {author} {\bibfnamefont {M.}~\bibnamefont {Noda}},
  \bibinfo {author} {\bibfnamefont {K.}~\bibnamefont {Nobusada}},\ and\
  \bibinfo {author} {\bibfnamefont {K.}~\bibnamefont {Yabana}},\ }\bibfield
  {title} {\bibinfo {title} {Time-dependent density functional theory for
  interaction of ultrashort light pulse with thin materials},\ }\href@noop {}
  {\bibfield  {journal} {\bibinfo  {journal} {Physical Review B}\ }\textbf
  {\bibinfo {volume} {98}},\ \bibinfo {pages} {245147} (\bibinfo {year}
  {2018})}\BibitemShut {NoStop}%
\bibitem [{\citenamefont {Yamada}\ and\ \citenamefont
  {Yabana}(2021)}]{yamada2021determining}%
  \BibitemOpen
  \bibfield  {author} {\bibinfo {author} {\bibfnamefont {S.}~\bibnamefont
  {Yamada}}\ and\ \bibinfo {author} {\bibfnamefont {K.}~\bibnamefont
  {Yabana}},\ }\bibfield  {title} {\bibinfo {title} {Determining the optimum
  thickness for high harmonic generation from nanoscale thin films: An ab
  initio computational study},\ }\href@noop {} {\bibfield  {journal} {\bibinfo
  {journal} {Physical Review B}\ }\textbf {\bibinfo {volume} {103}},\ \bibinfo
  {pages} {155426} (\bibinfo {year} {2021})}\BibitemShut {NoStop}%
\bibitem [{\citenamefont {Noda}\ \emph {et~al.}(2019)\citenamefont {Noda},
  \citenamefont {Sato}, \citenamefont {Hirokawa}, \citenamefont {Uemoto},
  \citenamefont {Takeuchi}, \citenamefont {Yamada}, \citenamefont {Yamada},
  \citenamefont {Shinohara}, \citenamefont {Yamaguchi}, \citenamefont {Iida}
  \emph {et~al.}}]{noda2019salmon}%
  \BibitemOpen
  \bibfield  {author} {\bibinfo {author} {\bibfnamefont {M.}~\bibnamefont
  {Noda}}, \bibinfo {author} {\bibfnamefont {S.~A.}\ \bibnamefont {Sato}},
  \bibinfo {author} {\bibfnamefont {Y.}~\bibnamefont {Hirokawa}}, \bibinfo
  {author} {\bibfnamefont {M.}~\bibnamefont {Uemoto}}, \bibinfo {author}
  {\bibfnamefont {T.}~\bibnamefont {Takeuchi}}, \bibinfo {author}
  {\bibfnamefont {S.}~\bibnamefont {Yamada}}, \bibinfo {author} {\bibfnamefont
  {A.}~\bibnamefont {Yamada}}, \bibinfo {author} {\bibfnamefont
  {Y.}~\bibnamefont {Shinohara}}, \bibinfo {author} {\bibfnamefont
  {M.}~\bibnamefont {Yamaguchi}}, \bibinfo {author} {\bibfnamefont
  {K.}~\bibnamefont {Iida}}, \emph {et~al.},\ }\bibfield  {title} {\bibinfo
  {title} {Salmon: Scalable ab-initio light--matter simulator for optics and
  nanoscience},\ }\href@noop {} {\bibfield  {journal} {\bibinfo  {journal}
  {Computer Physics Communications}\ }\textbf {\bibinfo {volume} {235}},\
  \bibinfo {pages} {356} (\bibinfo {year} {2019})}\BibitemShut {NoStop}%
\bibitem [{Sal()}]{Salmon:Online}%
  \BibitemOpen
  \href@noop {} {\bibinfo {title} {Salmon official}},\ \bibinfo {howpublished}
  {\url{http://salmon-tddft.jp}}\BibitemShut {NoStop}%
\bibitem [{\citenamefont {Perdew}\ and\ \citenamefont
  {Wang}(1992)}]{perdew1992accurate}%
  \BibitemOpen
  \bibfield  {author} {\bibinfo {author} {\bibfnamefont {J.~P.}\ \bibnamefont
  {Perdew}}\ and\ \bibinfo {author} {\bibfnamefont {Y.}~\bibnamefont {Wang}},\
  }\bibfield  {title} {\bibinfo {title} {Accurate and simple analytic
  representation of the electron-gas correlation energy},\ }\href@noop {}
  {\bibfield  {journal} {\bibinfo  {journal} {Physical review B}\ }\textbf
  {\bibinfo {volume} {45}},\ \bibinfo {pages} {13244} (\bibinfo {year}
  {1992})}\BibitemShut {NoStop}%
\bibitem [{\citenamefont {Von~Barth}\ and\ \citenamefont
  {Hedin}(1972)}]{von1972local}%
  \BibitemOpen
  \bibfield  {author} {\bibinfo {author} {\bibfnamefont {U.}~\bibnamefont
  {Von~Barth}}\ and\ \bibinfo {author} {\bibfnamefont {L.}~\bibnamefont
  {Hedin}},\ }\bibfield  {title} {\bibinfo {title} {A local
  exchange-correlation potential for the spin polarized case. i},\ }\href@noop
  {} {\bibfield  {journal} {\bibinfo  {journal} {Journal of Physics C: Solid
  State Physics}\ }\textbf {\bibinfo {volume} {5}},\ \bibinfo {pages} {1629}
  (\bibinfo {year} {1972})}\BibitemShut {NoStop}%
\bibitem [{\citenamefont {Oda}\ \emph {et~al.}(1998)\citenamefont {Oda},
  \citenamefont {Pasquarello},\ and\ \citenamefont {Car}}]{oda1998fully}%
  \BibitemOpen
  \bibfield  {author} {\bibinfo {author} {\bibfnamefont {T.}~\bibnamefont
  {Oda}}, \bibinfo {author} {\bibfnamefont {A.}~\bibnamefont {Pasquarello}},\
  and\ \bibinfo {author} {\bibfnamefont {R.}~\bibnamefont {Car}},\ }\bibfield
  {title} {\bibinfo {title} {Fully unconstrained approach to noncollinear
  magnetism: application to small fe clusters},\ }\href@noop {} {\bibfield
  {journal} {\bibinfo  {journal} {Physical review letters}\ }\textbf {\bibinfo
  {volume} {80}},\ \bibinfo {pages} {3622} (\bibinfo {year}
  {1998})}\BibitemShut {NoStop}%
\bibitem [{\citenamefont {Morrison}\ \emph {et~al.}(1993)\citenamefont
  {Morrison}, \citenamefont {Bylander},\ and\ \citenamefont
  {Kleinman}}]{morrison1993nonlocal}%
  \BibitemOpen
  \bibfield  {author} {\bibinfo {author} {\bibfnamefont {I.}~\bibnamefont
  {Morrison}}, \bibinfo {author} {\bibfnamefont {D.}~\bibnamefont {Bylander}},\
  and\ \bibinfo {author} {\bibfnamefont {L.}~\bibnamefont {Kleinman}},\
  }\bibfield  {title} {\bibinfo {title} {Nonlocal hermitian norm-conserving
  vanderbilt pseudopotential},\ }\href@noop {} {\bibfield  {journal} {\bibinfo
  {journal} {Physical Review B}\ }\textbf {\bibinfo {volume} {47}},\ \bibinfo
  {pages} {6728} (\bibinfo {year} {1993})}\BibitemShut {NoStop}%
\bibitem [{\citenamefont {Liu}\ \emph {et~al.}(2017)\citenamefont {Liu},
  \citenamefont {Li}, \citenamefont {You}, \citenamefont {Ghimire},
  \citenamefont {Heinz},\ and\ \citenamefont {Reis}}]{liu2017high}%
  \BibitemOpen
  \bibfield  {author} {\bibinfo {author} {\bibfnamefont {H.}~\bibnamefont
  {Liu}}, \bibinfo {author} {\bibfnamefont {Y.}~\bibnamefont {Li}}, \bibinfo
  {author} {\bibfnamefont {Y.~S.}\ \bibnamefont {You}}, \bibinfo {author}
  {\bibfnamefont {S.}~\bibnamefont {Ghimire}}, \bibinfo {author} {\bibfnamefont
  {T.~F.}\ \bibnamefont {Heinz}},\ and\ \bibinfo {author} {\bibfnamefont
  {D.~A.}\ \bibnamefont {Reis}},\ }\bibfield  {title} {\bibinfo {title}
  {High-harmonic generation from an atomically thin semiconductor},\
  }\href@noop {} {\bibfield  {journal} {\bibinfo  {journal} {Nature Physics}\
  }\textbf {\bibinfo {volume} {13}},\ \bibinfo {pages} {262} (\bibinfo {year}
  {2017})}\BibitemShut {NoStop}%
\bibitem [{\citenamefont {Yoshikawa}\ \emph {et~al.}(2019)\citenamefont
  {Yoshikawa}, \citenamefont {Nagai}, \citenamefont {Uchida}, \citenamefont
  {Takaguchi}, \citenamefont {Sasaki}, \citenamefont {Miyata},\ and\
  \citenamefont {Tanaka}}]{yoshikawa2019interband}%
  \BibitemOpen
  \bibfield  {author} {\bibinfo {author} {\bibfnamefont {N.}~\bibnamefont
  {Yoshikawa}}, \bibinfo {author} {\bibfnamefont {K.}~\bibnamefont {Nagai}},
  \bibinfo {author} {\bibfnamefont {K.}~\bibnamefont {Uchida}}, \bibinfo
  {author} {\bibfnamefont {Y.}~\bibnamefont {Takaguchi}}, \bibinfo {author}
  {\bibfnamefont {S.}~\bibnamefont {Sasaki}}, \bibinfo {author} {\bibfnamefont
  {Y.}~\bibnamefont {Miyata}},\ and\ \bibinfo {author} {\bibfnamefont
  {K.}~\bibnamefont {Tanaka}},\ }\bibfield  {title} {\bibinfo {title}
  {Interband resonant high-harmonic generation by valley polarized
  electron--hole pairs},\ }\href@noop {} {\bibfield  {journal} {\bibinfo
  {journal} {Nature communications}\ }\textbf {\bibinfo {volume} {10}},\
  \bibinfo {pages} {1} (\bibinfo {year} {2019})}\BibitemShut {NoStop}%
\bibitem [{\citenamefont {Ohmura}\ \emph {et~al.}(2004)\citenamefont {Ohmura},
  \citenamefont {Nakanaga},\ and\ \citenamefont {Tachiya}}]{Ohmura2004}%
  \BibitemOpen
  \bibfield  {author} {\bibinfo {author} {\bibfnamefont {H.}~\bibnamefont
  {Ohmura}}, \bibinfo {author} {\bibfnamefont {T.}~\bibnamefont {Nakanaga}},\
  and\ \bibinfo {author} {\bibfnamefont {M.}~\bibnamefont {Tachiya}},\
  }\bibfield  {title} {\bibinfo {title} {Coherent control of photofragment
  separation in the dissociative ionization of ibr},\ }\href
  {https://doi.org/10.1103/PhysRevLett.92.113002} {\bibfield  {journal}
  {\bibinfo  {journal} {Phys. Rev. Lett.}\ }\textbf {\bibinfo {volume} {92}},\
  \bibinfo {pages} {113002} (\bibinfo {year} {2004})}\BibitemShut {NoStop}%
\end{thebibliography}%
\eject 
\end{document}